%% file: main.tex
\documentclass[journal]{IEEEtran}
\usepackage[utf8]{inputenc}
\usepackage[english]{babel}
\usepackage{subfiles}
\usepackage[pdftex]{graphicx}
\usepackage{subfig}
\usepackage{tabularx}
\usepackage{multirow}
\usepackage{dblfloatfix} 
\usepackage{hyperref}
\usepackage{enumitem}

\usepackage{amsmath}
\usepackage{color,soul}
\usepackage{algorithm,algpseudocode,float}

\usepackage{cite}

\ifCLASSINFOpdf
\else
\fi

\usepackage{amsmath}

\usepackage{array}
\usepackage{tabularx}
\usepackage{multirow}
\usepackage{multicol}

\begin{document}
%
\title{PULSE: Optical Circuit Switched Data Center Architecture Operating at Nanosecond Timescales}

\def\UrlBreaks{\do\/\do-}
\author{Joshua~L.~Benjamin,~\IEEEmembership{Student Member,~IEEE},
Thomas Gerard,~\IEEEmembership{Student Member,~IEEE},
        Domani\c{c}~Lavery,~\IEEEmembership{Member,~IEEE},~Polina~Bayvel,~\IEEEmembership{Fellow,~IEEE}
        and~Georgios~Zervas,~\IEEEmembership{Member,~IEEE}
        \vspace{-15pt}
\thanks{The authors are with the Department
of EEE, University College London, United Kingdom (E-mail: g.zervas@ucl.ac.uk)}}

\maketitle

\begin{abstract}
We introduce PULSE, a sub-$\mu$s optical circuit switched data centre network architecture controlled by distributed hardware schedulers. PULSE is a flat architecture that uses parallel passive coupler-based broadcast and select networks. We employ a novel transceiver architecture, for dynamic wavelength-timeslot selection, to achieve a reconfiguration time down to O(100ps), establishing timeslots of O(10ns). A novel scheduling algorithm that has a clock period of 2.3ns performs multiple iterations to maximize throughput, wavelength usage and reduce latency, enhancing the overall performance. In order to scale, the single-hop PULSE architecture uses sub-networks that are disjoint by using multiple transceivers for each \textcolor{black}{node} in 64 \textcolor{black}{node} racks. At the reconfiguration circuit duration (\textit{epoch} = 120~ns), the scheduling algorithm is shown to achieve up to 93\% throughput and 100\% wavelength usage of 64 wavelengths, incurring an average latency that ranges from 0.7-1.2~$\mu$s with best-case 0.4~$\mu$s median and 5~$\mu$s tail latency, limited by the timeslot (20~ns) and epoch size (120~ns). We show how the 4096-\textcolor{black}{node} PULSE architecture allows up to 260k optical channels to be re-used across sub-networks achieving a capacity of 25.6~Pbps with an energy consumption of 82 pJ/bit when using coherent receiver. 
\end{abstract}


\IEEEpeerreviewmaketitle

\subfile{sections/1.introduction}

\subfile{sections/2.OCS}

\subfile{sections/3.architecture}

\subfile{sections/4.scheduler}

\subfile{sections/5.results}

\subfile{sections/6.related}

\subfile{sections/7.conclusion}

\section*{Acknowledgment}
This work is supported by EPSRC TRANSNET program (EP/R035342/1), by Microsoft
Research through its PhD scholarship programme (T. Gerard) and the
UCL-Cambridge CDT program in
Integrated Photonic and Electronic Systems.

\ifCLASSOPTIONcaptionsoff
  \newpage
\fi

\bibliographystyle{IEEEtran}
\bibliography{reference}
%

\end{document}

%% file: sections/1.introduction.tex
\section{Introduction}

\IEEEPARstart{T}{he} rapid increase in the rate of intra-data center traffic, due to the growth of data services, requires the supportive growth of resources within data center networks (DCNs). Cisco’s Visual Networking Index (VNI) suggests that 73\% of DCN traffic is internal within the data center network \cite{cisco}. According to Google, the demand for bandwidth in data centers doubles every 12-15 months \cite{Google}. Intel anticipates that 70-80\% of their computing and storage systems will be deployed into data centers by 2025 \cite{Intel}. While data center operators are being forced to scale their computing resources they are also required to maintain low costs and power consumption. In 2015, total power consumption of DCNs worldwide was 416.2 terawatt hours, while the total power consumption in the UK was approximately 300 terawatt hours \cite{Independent}. In some cases, energy use is a substantial cost relative to the IT hardware itself \cite{APC}. Moreover, by 2021, about 95\% of all data center traffic will originate from the cloud \cite{cisco2}. In bursty, cloud based applications, 90\% of packets have a size of less than 576 bytes \cite{kari}; smaller packets require faster switching. However, current electronic packet switched networks have long tail latencies of several hundred milliseconds; orders of magnitude higher than median latency, degrading application performance \cite{amazon}. Hence, there is a need for networks, which reconfigure at nanosecond timescales and ensure low and deterministic tail latency and high network throughput. Hence, we propose the use of optically switched networks which can be tailored to showcase the aforementioned features.

Optical switch technologies, including Arrayed Waveguide Grating Routers (AWGRs) and star-coupler based switches, have been proven to scale to port counts as high as 512 or 1024 ports \cite{AWG,petabit}. With WDM, TDM and advanced optical modulation techniques, optical transceivers are also able to unlock and support higher data transmission rate (or bit rate) per port \cite{polina1}. However, a major challenge faced when scaling an optical switch is the scalability of the central scheduler \cite{Diluc}. Hence, many optical switch technologies resolve to software-based scheduling to simplify switch configuration \cite{AWG,petabit,griffin,potori}. Nevertheless, software based control systems take milliseconds to compute switch configurations. To tackle this, researchers have proposed optical switch solutions with distributed scheduling for simplified control. However, these proposals tend to elevate data plane complexity \cite{tonak,iris,mems}.  Recent optical packet switching technologies with fast switching control have also been shown to have scaling restrictions as they require complex optical data plane architectures, multiple optical components \cite{hippo} and packet management techniques to realize electronic packet switch functionalities \cite{OPS}. 

Previously, we proposed a transceiver-based optical circuit switched (OCS) network with a passive star-coupler core \cite{sigcomm}, scalable to 1000 ports \cite{adamjlt} that used a centralized scheduler to reconfigure the network by defining wavelength and timeslots at the transceivers \cite{jlb}. However, the broadcast and select network suffered from resource wastage as only $W$(=80) wavelengths were used for $N$(=1000) servers (network efficiency = 8\%) and had a long circuit duration (\textit{epoch}) of 2~$\mu$s, well above the aforementioned target of nanosecond timescales. In this paper, we propose PULSE, a novel scalable OCS architecture that supports nanosecond speed reconfiguration time, while enabling network and capacity scalability with the help of independent distributed hardware schedulers with high network efficiency ($W=N$). We introduce novel transceiver architectures that enable faster circuit reconfiguration time and evaluate the network energy consumption of the proposed transceiver combination. We introduce a novel scheduling algorithm that can effectively compute a new wavelength configuration (per \textcolor{black}{node}) for each timeslot within each \textit{epoch} (circuit cycle duration) and limit tail latency to a few microseconds. 

The PULSE architecture does not require in-network routing/switching, buffering and addressing. However, it requires ultra-fast (a) O(ns) scheduling, (b) tunable wavelength switching, (c) filtering, \textcolor{black}{(d) distributed/scalable transport network}, (e) clock and data recovery \cite{kari} and (f) O(100~ps) synchronization. Research is being carried out on several aspects of PULSE \textcolor{black}{but the primary focus of this paper is to:} 
\begin{itemize}

\item \textcolor{black}{propose a novel modular and distributed architecture (d) that eases control and data plane scalability (Section II).}

\item \textcolor{black}{propose novel ultra-fast O(ps-ns) wavelength-timeslot selective transceiver architectures (b-c) (Section III).} 

\item \textcolor{black}{propose a novel hardware scheduler design (a) and evaluate its scalability (Section IV).}

\item \textcolor{black}{review schedule performance under various epoch sizes, traffic distributions/loads to evaluate its effect on resource utilization, throughput, latency and transmitter/scheduler buffer size compared to previous work (Section V).} 

\item \textcolor{black}{study the implications of network and transceiver architecture on scalability, cost, power and latency overhead (Section V).} 
\end{itemize}

Section VI presents related work in the \textcolor{black}{optical circuit switching} field \textcolor{black}{in order} to identify the relevance and novelty of PULSE and we conclude in section VII.

%% file: sections/2.OCS.tex
\section{OCS Network Architecture}
\label{section_OCS}
\begin{center}
\begin{figure}[!ht]
\includegraphics[width=\linewidth]{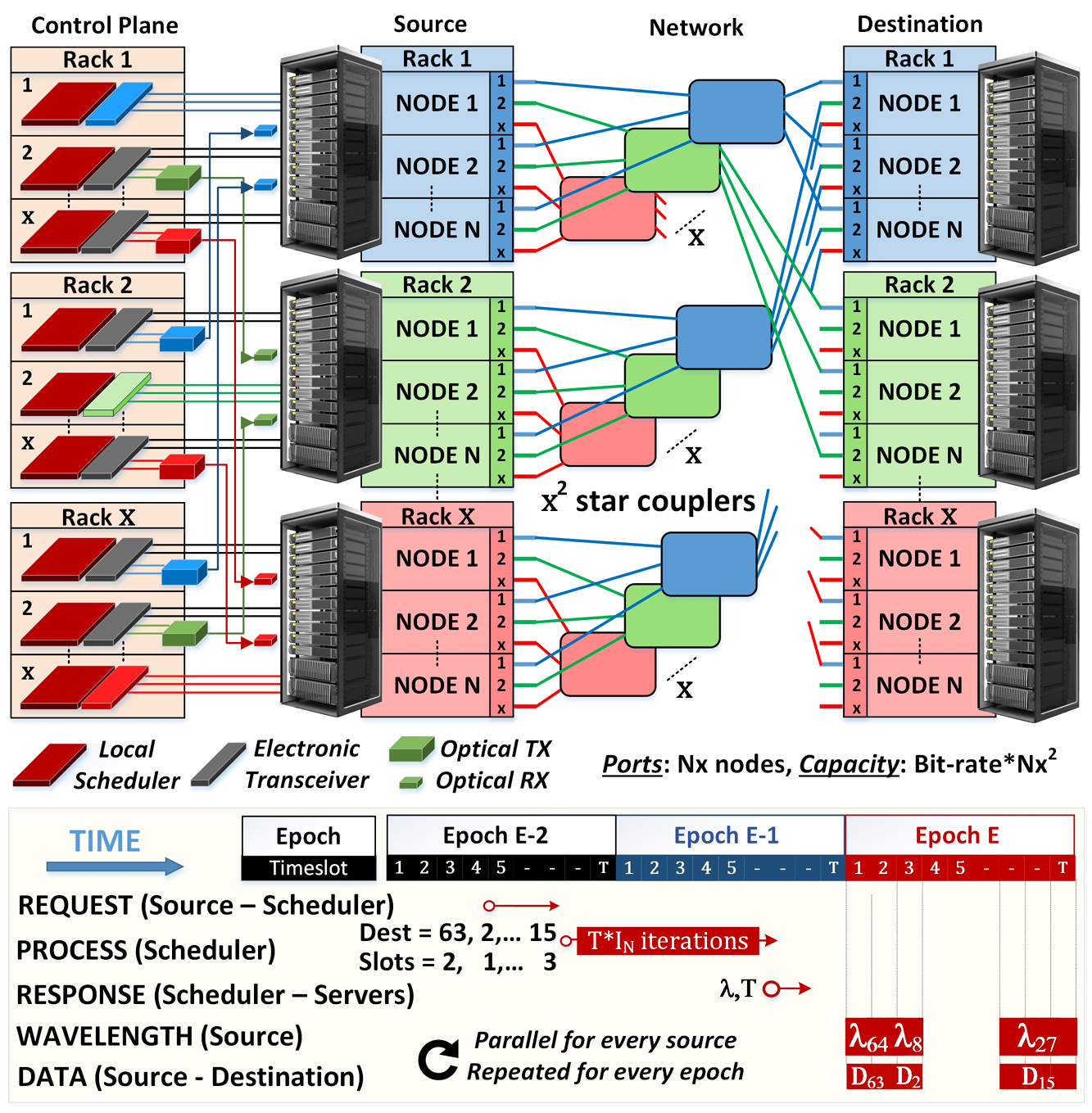}
\caption{\textcolor{black}{PULSE - Top: Parallel OCS network architecture with distributed hardware schedulers, Bottom: Control-Data handshake and dataflow.}}
\label{OCS} 
\vspace{-7pt}
\end{figure}
\end{center}
\vspace{-7pt}
The data plane of PULSE consists of parallel star-couplers grouped per rack, shown in the top right side of Fig.~\ref{OCS}. Each passive star-coupler core forms a broadcast and select network \cite{sigcomm}. There are up to $x$ racks, where each rack contains $N$ \textcolor{black}{nodes}. Each \textcolor{black}{node} houses up to $x$ optical transceivers, where each transceiver connects a \textcolor{black}{node} to a different star coupler and thereby, a different sub-network or rack. For connection establishment and eventual data transmission across the switch, the source transmitter and the destination receiver must both tune to the same wavelength and timeslot. As shown by the top left side of Fig. \ref{OCS}, for every star coupler network in the data plane, the control plane has a corresponding $N$-\textcolor{black}{node} scheduler co-hosted in the same rack, which processes the requests for that particular source-destination rack pair. The parallel OCS network proposed requires $x^2$ $N$-\textcolor{black}{node} star couplers and schedulers. The PULSE architecture scales to support up to $Nx$ \textcolor{black}{nodes} with a capacity of $BNx^2$, where  $B$ is the effective line-rate of each transceiver. 

As shown by bottom part of Fig. \ref{OCS}, the communication timeline groups timeslots to form \textit{epochs}. The scheduler also computes, for an epoch, wavelength-timeslot grants for a specific epoch. Figure \ref{OCS} shows a source \textcolor{black}{node} sending requests to the relevant scheduler that is associated with the destinations of interest ($D_{63}, D_{2}, D_{15}$) with the number of timeslots requested (2,1,3 respectively) and timeslot number well over an epoch in advance. The scheduler performs as many iterations as it can in one epoch to compute the wavelength ($\lambda_{64},\lambda_{8},\lambda_{27}$ in Fig. \ref{OCS}) and timeslots for each request. Table \ref{table_2} shows how the network, requests, racks, cables, channels and capacity scale while increasing the number of transceivers and racks ($x \in$ 4,8, 16, 32 and 64) at 64 \textcolor{black}{nodes} per rack ($N=64$).
\begin{table}[ht] 
\caption{PULSE: Scalability, Capacity, Complexity at $N=64$}
\centering
\label{table_2} 
\begin{tabular}{|l|c|c|c|c|c|}
\hline
\multicolumn{1}{|c|}{\multirow{2}{*}{\textbf{Network Parameters}}}  & \multicolumn{5}{c|}{\textbf{Transceivers per \textcolor{black}{node} (x)}} \\ \cline{2-6} 
\multicolumn{1}{|c|}{}                                    & \textbf{4}  & \textbf{8}  & \textbf{16}  & \textbf{32}  & \textbf{64} \\ \hline
\textbf{Total \textcolor{black}{nodes}}                           & 256 & 512 & 1024 & 2048 & 4096       \\ \hline
\textbf{Req/\textcolor{black}{node}/epoch ($R$)}                           & 24 & 48 & 96 & 192 & 384         \\ \hline
\textbf{Racks ($x$)}                         & 4          & 8          & 16  & 32 & 64       \\ \hline
\textbf{Sub-stars ($x^2$)}                   & 16          & 64          & 256 & 1024 & 4096         \\ \hline
\textbf{Cables ($N \times x^2 \times 4$)}                         & 4096          & 16384          & 65536 & 0.26M & 1.04M         \\ \hline
\textbf{Channels ($Wx^2$)}                             & 1024        & 4096        & 16384  &  65536 &  0.26M   \\ \hline
\textbf{Max capacity (Tbps)}                              & 100       & 400       & 1598 & 6394 & 25575     \\ \hline
\end{tabular}
\end{table}

The spatial division multiplexing (SDM) created by the parallel and distributed star-couplers enables the re-use of wavelengths. In other words, the same wavelength can be re-used in another star-coupler. The complete independent nature of individual sub-networks means that local schedulers have no dependency on the traffic or resource usage faced by other stars. The synchronization problem is also reduced to a local sub-network and not required at a global level.  Each 64-port sub-star uses 64 wavelengths and the SDM enables 4096 parallel uses of wavelengths allowing a total of 0.26M channels and a network with an enhanced capacity of {\raise.17ex\hbox{$\scriptstyle\mathtt{\sim}$}} 25.6 Pbps, accounting for tuning overhead. \textcolor{black}{Each sub-network creates shareable bandwidth resources between two racks. The overall network bandwidth (between all racks) is indicated by `Max capacity' in Table \ref{table_1}. However, at any instant, the maximum node-to-node capacity is 100 Gbps.} 

\textcolor{black}{As shown in table \ref{table_1}, each node uses up to 6.4~Tbps. PULSE is a network solution, where each node can be a high-performance computational resource in a heterogeneous cloud DCNs - CPU, GPU, TPU, HBM ($>$1~Tbps) or ToRs. Resources like HBMs and GPUs \cite{gpu_cap} require more than 1 Tbps bandwidth. Large multi-core chips are also being explored to support fast AI calculations \cite{ray_2019}.  There has been considerable growth in the performance of GPUs, nearly 1.5 times a year \cite{gpu_growth} and reaching 1.5~Tbps by 2020 \cite{gpu_cap}. Although network and computational processing are capable of handling high throughput (100~Tbps), NIC or data ingest capacity constraints (100~Gbps) create bottlenecks, leading to inefficient systems that constraint applications to operate locally and degrade the overall application performance. Hence, the FastNICs program initiated by DARPA \cite{insidehpc_2019} aims to boost network NIC and stack capacity by 100 times to accelerate distributed application performance and close the gap between processing and network capability. Hence, it is expected that in the future end-nodes will support high bandwidths (6.4 Tbps). Today’s expensive network switches are over-designed and under-utilized \cite{stardust}. PULSE intends to maximize bandwidth utilization even if high capacities are generated at the nodes. }

%% file: sections/3.architecture.tex
\section{Optical Circuit Switch Elements}

\subsection{Transceiver technology}
\begin{figure}[!t]
\centering
\includegraphics[width=1\linewidth]{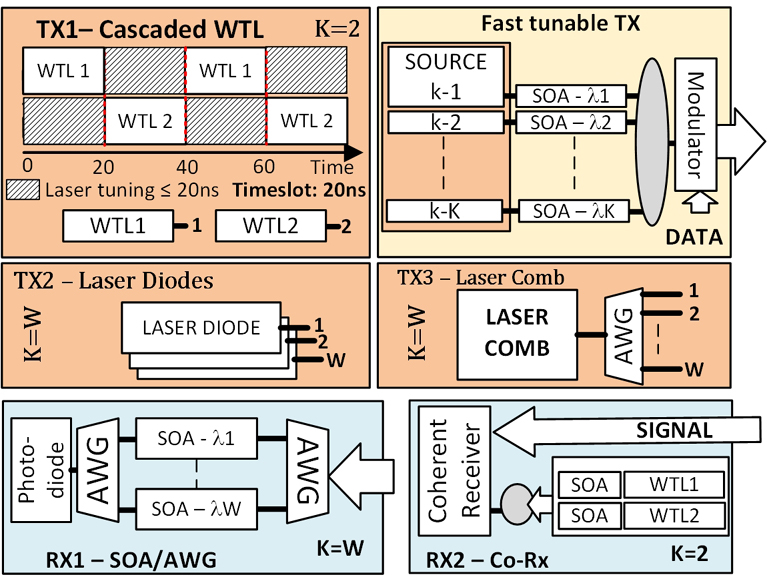}
\caption{\textcolor{black}{Transceiver options for PULSE: TX1-Cascaded WTL, TX2-Laser Diodes, TX3-Laser Comb, RX1-SOA/AWG, RX2-Coherent Rx.}}
\label{SOA} 
\vspace{-7pt}
\end{figure}

In this work, we propose the use of \textcolor{black}{fast} wavelength \textcolor{black}{and timeslot }selectable transceivers, \textcolor{black}{as shown in Fig. \ref{SOA}. For the purpose of evaluating latency, cost, power and throughput, we assume a line-rate of 100~Gbps per wavelength. The line-rate could be achieved by adopting PAM-4 (56~Gbaud) for direct-detect and DP-QPSK (4$\times$28Gbaud) for coherent systems; though the wavelength-timeslot switching is format/rate flexible.} At 100 Gbps, this system targets the transfer of 250~bytes per 20~ns time-slot, which corresponds to the overall median packet size across various data center workflows \cite{packetsize}. As the globally synchronized optical switch only needs  to carry the bare payload of the packet, the effective packet length is slightly more than an average Ethernet packet. In addition, prior work has experimentally demonstrated fast time-slot switching with a 1-bit guard band, minimizing any TDM overhead \cite{adamofc}. At each transmitter, smaller packets ($\ll$250 bytes) to the same destination are aggregated for effective slot usage. 

\begin{figure}[!b]

\begin{center}
\includegraphics[width=0.8\linewidth]{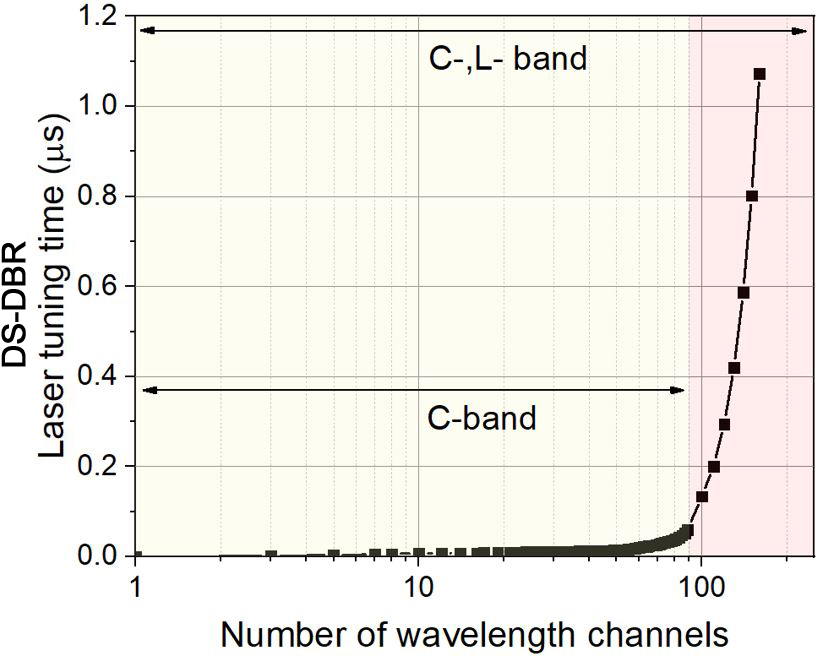}
    
\caption{DS-DBR laser tuning time prediction versus the number of available wavelength channels with 99.9\% regression on \cite{adamjlt}.}
\label{fig_laser} 
\end{center}
\end{figure}

\subsubsection{Prior work}
Dynamic wavelength assignment (using software algorithms) with all processing and buffering at the edge was proposed in \cite{polina1, polina2}. The work in \cite{adamofc, adamjlt} proposed each node to be equipped with a tunable DS-DBR laser and a coherent receiver with an independently tunable DS-DBR local oscillator laser. This enables fast wavelength selectivity, O(10ns), at both the transmitter and receiver, and the high sensitivity of the coherent receiver also enables scalability in the data plane since it allows for a larger system loss budget and thus a higher port count star-coupler \cite{adamofc,adamjlt}. Although a 1000-port star coupler network can be created, experiments have shown that only up to 89 wavelengths can be supported by the laser hardware positioned on a 50 GHz ITU grid within the optical C-band to ensure minimal crosstalk between channels ($W\ll{}N$). Extending the same grid to the L-band could allow the use of 160 wavelengths. However, Fig. \ref{fig_laser} shows the growth of this tuning time when using an extrapolated regression model with greater than 99.9\% confidence interval; this predicts that a 160-wavelength system takes more than 1$\mu$s to tune all transceiver pairs. The large tuning time has a direct impact on switch latency and tail latencies, which degrades network performance, setting a limit on the wavelength channels allowed, assuming the tunable DS-DBR lasers are used at the transceivers. This performance degradation has an impact on the scalability of the OCS network in terms of capacity, limiting $W\ll{}N$. Regardless of the number of wavelength channels used, the large tuning time would require a dedicated synchronized tuning time prior to every epoch, which lasted in O(1$\mu$s) to minimize overhead.

\subsubsection{Transmitter options}
Following the previous argument, we explore methods of approaching a larger number of wavelength channels ($W$), while making sure that optical transmitters do not impede switching times. With that aim, we consider 3 WDM transmitter architectures that have the potential to achieve sub-nanosecond switching. Electro-optical amplifier switch based on chip-on-carrier SOA with tunable lasers were experimentally demonstrated to achieve switching times of up to 115 ps by \cite{soa100}. Here, we use $k$ such SOAs at the transceiver, in order to enable reconfiguration at a faster rate of the O(10ns), connected to one of three laser source options that we propose as shown in Fig. \ref{SOA}. In our transceiver proposal, the first transmitter laser source option (TX1) is to employ $k$ \textcolor{black}{widely tunable lasers (WTLs). The WTL could be the DS-DBR lasers described in the previous sub-section or SG-DBR lasers} in a cascaded fashion, connected to $k$ SOA gates, as shown in Fig. \ref{SOA}. \textcolor{black}{However, the WTL is required to tune to a new wavelength within 20~ns. Modulated grating Y-branch (MG-Y) lasers \cite{finis} can be used as WTLs to achieve fast wavelength selection within 20~ns. The work in \cite{five_nano} has shown SG-DBRs to achieve wavelength tuning in 5~ns. Prior experiments have also shown that 67\% of transitions between any DS-DBR equipped transceiver pair of the 89 available wavelength channels (59 channels) complete tuning within 20 ns \cite{adamjlt}; work is underway to improve this to support 64 channels. As shown in Fig. \ref{SOA} (TX1), $k=2$ cascaded WTLs (and SOAs) are required at the transmitter and receiver to create 20~ns timeslots.} This will further reduce cost and power consumption. \textcolor{black}{The star-coupler core has a loss of $-3log_2N$~dB, which corresponds to -18 dB for a 64-port network. The SOAs used at the transmitter (and receiver option 1) compensate for this loss by providing a +10 to +15~dB amplification \cite{soa100} (per SOA per path). Coherent receiver (receiver option 2) are highly sensitive (-21~dB with BER of $10^{12}$) \cite{adamjlt} and can accommodate a larger split.} In this paper, we aim to increase the network transceiver efficiency to 100\% by increasing the number of channels to the number of nodes in the sub-network ($W=N$). Hence, the paper will investigate the network performance for ($N$=) 64-node racks and ($x$=)16 racks, scaling to 1024 nodes. The second transmitter laser source option (TX2) is the use of $k$(=W) VCSEL laser diodes \cite{vcsel}, one for each wavelength, at the source connected to $k$ SOA gates as shown in Fig \ref{SOA}. The third transmitter laser source option (TX3) is to use a laser comb that generates $k$(=W) wavelength sources connected $k$ SOA gates \cite{comb}.

\subsubsection{Receiver options}
We propose two options for the receivers, as shown at the bottom of Fig. \ref{SOA}. The first receiver option (RX1) contains an array of $W$ SOAs surrounded by AWGRs, followed by a direct detection photodiode. The selection of the SOA allows the fast wavelength selection (O(100~ps)) at the receiver. The disadvantage of such a receiver is the requirement of many SOA gates, which has an impact on overall power consumption. The second receiver option (RX2) contains $k$(=3) DS-DBR tunable transmitters with SOA gates, as at the transmitter, which serve as local oscillators (LOs) for a coherent receiver. 

\subsection{Control Plane}
Each node sends requests to the scheduler a \textit{few} epochs in advance (depending on the dominant propagation delay) and awaits response before reconfiguring transceiver wavelengths for the subsequent epoch. A control sub-network requires all $N$ nodes to be connected to the local scheduler. Requests sent from each of the nodes have the following structure: requested destination (6 bits), slot size (up to 5 bits for 600~ns epoch) and epoch stamps (\textcolor{black}{8} bits) to identify the requests. Once received, the central scheduler stores requests from all nodes in a 1.2~kB buffer, which can store up to $R$(=6 requests) per node per epoch. The control request communication needs to communicate \textcolor{black}{19} bits within a timeslot (20~ns), and hence, requiring a \textcolor{black}{1} Gbps link. While running up to $I$ iterations to process the requests, the scheduler stores the wavelength-timeslot pair grant information in a second buffer of 1.2~kB. Once the schedule is computed for an entire epoch, the wavelength-timeslot pair \textcolor{black}{grant information is communicated. Each grant has the following structure: TX/RX wavelength (6 bits each), destination node (6 bits), valid bit (1 bit) 19~bits for 64-port network. Hence, the grant communication also requires a 1~Gbps transceiver link to send 19 bits in every timeslot (20~ns).} As shown in Fig. \ref{OCS}, the control plane of each sub-network is co-hosted within every source rack to keep the request-response handshake propagation delay to a known minimal constant ($\approx 30$~ns). Regardless of where the destination rack is hosted, the control plane propagation delay is a constant, as will be discussed further in section V.

%% file: sections/4.scheduler.tex
\begin{figure*}[!t]
    \includegraphics[width=1\textwidth]{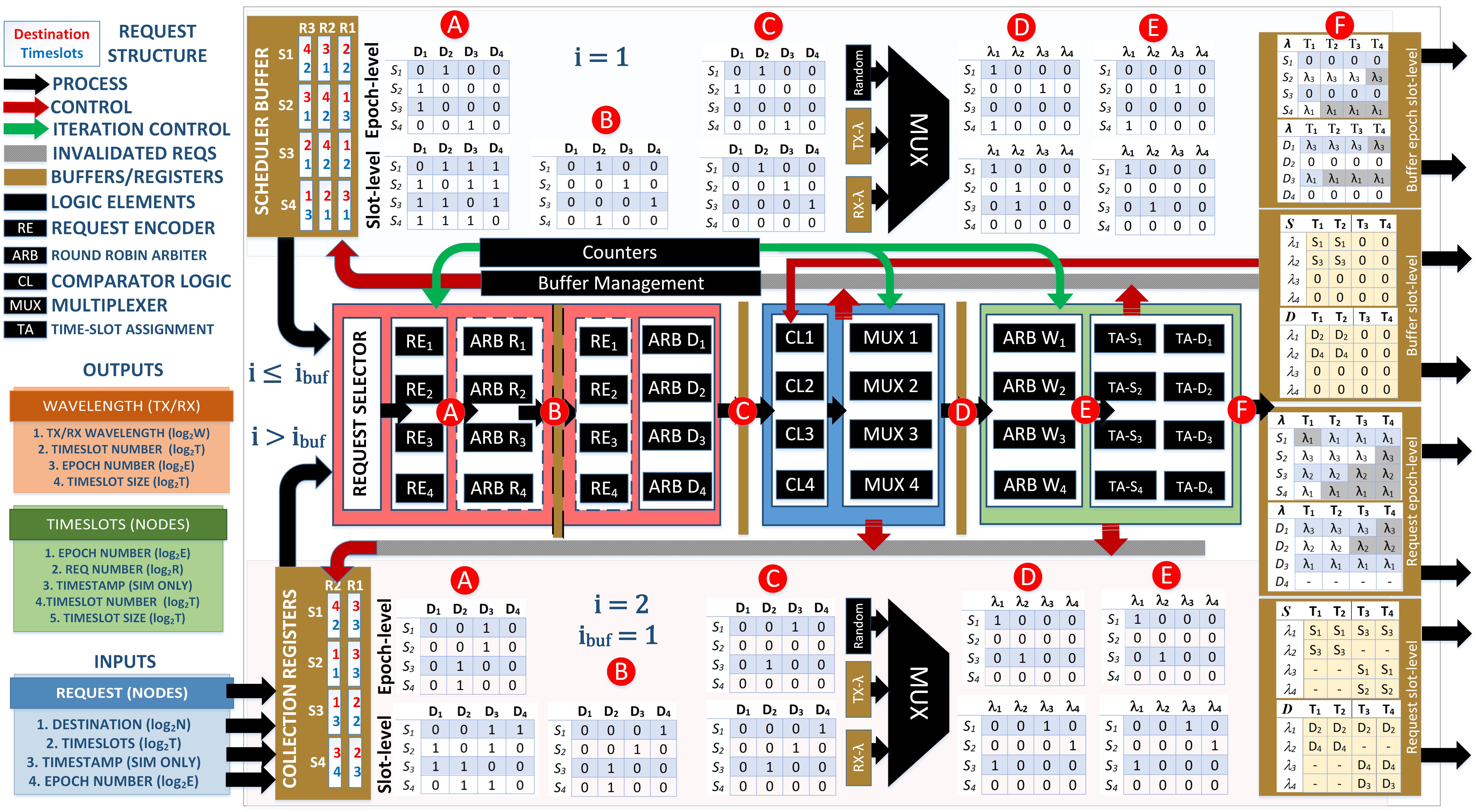}
    \caption{Epoch/Slot-level algorithm: 4-port central scheduler showcasing three hardware stages with a 4-port scheduling example showing two iterations, first handling requests from buffer (top), then from the \textcolor{black}{node} (bottom) (when $i_{buf}=1$).}
    \label{fig_SCHD}
\vspace{-7pt}
\end{figure*}

\section{Hardware Scheduler Algorithms}
In this section, we describe two hardware implementable scheduling algorithms, which can use any of the fast tunable transceiver options shown in Fig. \ref{SOA}. To meet the strict timing requirements, we adopt a parallel design that considers at least $N$ requests, one from each of the $N$ \textcolor{black}{nodes}. Although parallelism speeds up the scheduling process, it also creates contention for wavelength and timeslot resources. We use round-robin arbiters in our hardware design to ensure the fair selection of up to $N$ contention free source-destination requests, with unique source and destination ports, per clock cycle. The selected contention-free node-pairs are assigned wavelengths (WDM) and timeslots (TDM) in parallel. The scheduler aims to perform $I$ iterations within one epoch to maximize throughput. The number of iterations that the scheduling algorithm can perform is limited by the epoch size ($E \in 120, 360, 600$ns) and the clock period of the hardware ($clk$=2.3~ns for 64-port scheduler). Requiring $b$ cycles to boot up ($b=3$ for epoch-level scheduling, $b=4$ for slot-level scheduling), the maximum number of scheduler iterations is  $I=\left\lfloor{}E/\text{clk}\right\rfloor-b$. After $I$ iterations, failed requests are buffered and are given a chance to retry in subsequent epochs. 

We propose two distinct algorithms for resource (wavelength and timeslot) allocation: epoch-level and slot-level scheduling algorithms. As the name suggests, the epoch-level scheduling algorithm aims to tune the transceiver wavelengths \textit{once} at the beginning of every epoch. However, in slot-level scheduling algorithm, wavelength is tuned \textit{once} between each timeslot. The tuning overhead is higher for slot-level scheduling and it is taken into account while discussing throughput results in section V. From a scheduling perspective, the slot-level scheduling algorithm unlocks a new level of flexibility by reducing contradictions created by parallel wavelength assignments. Although both the epoch-level and slot-level scheduling algorithms use similar hardware elements, the major difference between them lies in the strategy used to allocate resources, which will be discussed in this section.

\subsection{Hardware Scheduler Design}
The hardware logical elements used to synthesise and implement the scheduling algorithms are discussed in sub-section. The dominant element used in the realization of the scheduling algorithm is the round-robin arbiter shown in Fig. \ref{fig_SCHD}, as they are scalable \cite{jlb}. The critical path lies in the carry chain of the arbiters, like in digital adders. Using optimal length carry look-ahead generators can provide high-speed arbitration by shortening the critical path. We also show how the logical hardware components work together to efficiently allocate resources by using an example showcased in Fig. \ref{fig_SCHD}. Figure \ref{fig_SCHD} shows the epoch-level and slot-level scheduling for a 4-port scheduler and the iterations dealing with buffer ($i$=1 shown by the top half) and dealing with requests from \textcolor{black}{node} ($i$=2 shown by the bottom half). 

\subsubsection{Node Contention Resolution (NCR)}
This module is shown by the red block in Fig. \ref{fig_SCHD}. In epoch-level scheduling, the NCR module has only one pipeline stage containing $N$ parallel $N$-bit request encoders (RE) and $N$ parallel round-robin arbiters (ARB~D). The hardware elements in Fig. \ref{fig_SCHD} that have dotted outline do not exist in the epoch-level scheduling, but only exist in the slot-level scheduling. In epoch level scheduling, every clock cycle deals with one request per source. Each RE converts the destination requested into a one-hot binary vector to create a compatible input for the round-robin arbiter that follows. The matrices labelled as epoch-level in Fig. \ref{fig_SCHD} and denoted by `A' show the encoding of the \textit{first} requests from each source ($R_1$), both in the buffer iteration (epoch-level matrix at the top of figure, $i=1$) and the \textcolor{black}{node} requests (epoch-level matrix at the bottom of figure, $i=2$), to form $N \times N$-bit matrices. Each of the $N$ round-robin arbiters deals with resolving contention for one destination (ARB~D). All $N$ sources for a particular destination are given as an input to the destination round-robin arbiters and only one request is successful per destination per iteration (labelled as epoch-level in Fig. \ref{fig_SCHD}, denoted by `C'). 

In contrast, the NCR module in the slot-level scheduling algorithm has two pipeline stages. The first stage has $N$ parallel $N$-bit RE and $N$ \textit{source} round-robin arbiters (ARB~R). The second stage, containing the same hardware components as the first stage, has $N$ parallel $N$-bit RE and $N$ \textit{destination} round-robin arbiters (ARB~D). Notice that the matrices formed after encoding in Fig. \ref{fig_SCHD}, denoted by A, contains all destination requests ($R1, R2$ and $R3$). This is the case in both the requests from the buffer (slot-level matrix at the top of figure, $i$=1) and the new requests from the \textcolor{black}{nodes} (slot-level matrix at the bottom of figure, $i$=2). Considering multiple requests from the same source, the slot-level scheduler requires both source arbiters, ARB~R (source contentions are resolved in slot-level matrix `B'), and destination arbiters, ARB~D (destination contentions are resolved in slot-level matrix `C'). The pipelining of arbiters in slot-level scheduling, however, creates repeated grants. This is dealt with using a feed-forward control; if requests are repeated or have already been successful in previous iteration they are cancelled in the second stage. 

Although Fig. \ref{fig_SCHD} does not show a decoder, each pipeline stage in the NCR module contains a decoder in both scheduling algorithms. The decoders help to minimize the register size after each stage to $log_2N$ bits per source to store only the winning destination, instead of $N$ bits per source.

\subsubsection{Wavelength Decision (WD)}
The wavelength decision module (WD), shown by the blue block in Fig. \ref{fig_SCHD}, is a pipeline stage that has the same hardware and functionality in both epoch-level and slot-level scheduling algorithms. This module/stage is composed of a parallel comparator logic (CL) blocks and parallel multiplexers (MUX), one per source, and the select signal of the MUX is tied to the resource registers at the last stage. This stage works on contention-free node-pairs and uses information from registers to perform a parallel check on node-pairs, by reading previous resource allocations of both the source and the destination of interest in a particular iteration. This is shown by the red line feeding back from registers to CL in Fig. \ref{fig_SCHD}. If contending wavelengths have already been assigned to the node pair of interest, the request is invalidated for the current epoch (shown by the red arrow going to the shaded gray lines going to request collect registers to buffer registers). If either only the transmitter or only the receiver has been assigned a wavelength, then that particular wavelength is selected (shown by the MUX in the diagram), subject to  time-slot availability. If there is no wavelength assignment history for the transmitter and receiver and resources are available, a random wavelength is assigned, subject to  time-slot availability. In the hardware, a ROM is implemented to hold $N$ arrays of $T \times log_2W$ pseudo-random wavelengths (5-25 bytes per source). The selection of wavelengths in this stage does not guarantee a successful grant. The selection of same wavelengths in parallel can create contention in the parallel resource allocation process. Hence, the selected wavelength assignments and the node-pairs are stored in registers and forwarded to the next stage. Again, a decoder is used in both epoch-level and slot-level scheduling algorithm to store the wavelength decision in $log_2W$ bits per source-destination pair (in addition to the $log_2N$ bits for destination).

\subsubsection{Resource allocation or Wavelength Contention Resolution (WCR)}
The third module, wavelength contention resolution (WCR), is shown by the green block shown in Fig. \ref{fig_SCHD}. Both epoch-level and slot-level scheduling algorithms contain this pipeline stage. Firstly, this stage contains request encoders (not shown in figure), which make $N$ parallel $W$-bit one-hot binary vector that translate the wavelength decisions into compatible inputs for $N$ parallel round-robin arbiters. The formation of this source(-destination pair) and wavelength matrix is denoted as `D' in Fig. \ref{fig_SCHD} for both epoch-level and slot-level scheduling algorithms. Secondly, this module uses $W$ parallel $N$-bit round robin arbiters (ARB~W) to resolve contention between wavelength assignments (`E' in Fig. \ref{fig_SCHD}). Up to $W$ winning grants are generated in parallel by the arbiters. Thirdly, this module contains timeslot allocators for both the source (TA-S) and destination (TA-D). 

The encoders and arbiters have the same functionality in both algorithms. However, the epoch-level and slot-level scheduling work differently in the timeslot allocation blocks. In epoch-level scheduling, the winning grants of the arbiters are granted as many time-slots as they have requested, subject to availability. If only fewer slots are available, then available slots are granted and the request is updated with the new slot size and buffered (shown by the red line to the buffer in Fig. \ref{fig_SCHD}) for processing in subsequent epochs. The final allocations in epoch-level scheduling for both for buffer ($i$=1-top) and \textcolor{black}{node} ($i$=2-bottom) requests and for both the source and the destination, are shown by the matrices denoted by `F' in Fig. \ref{fig_SCHD}. In epoch-level scheduling a wavelength allocation means that the source must use the same wavelength for all timeslots in the epoch. Hence, in the epoch-level matrices in `F', the timeslots that are greyed out show that the timeslot is unused, but the wavelength for that particular source/destination is locked for the entire epoch. 

In slot-level scheduling, a wavelength allocation only means that the wavelength is locked for that particular timeslot. Hence, a different allocation strategy is required. The slot-level scheduling algorithm deals with this by having two iteration phases: coarse allocation and fine allocation. The first few iterations, based on $T/S_{avg}$ (number of timeslots in epoch/average slot size requested) are handled with coarse allocation, where parallel requests are allocated as many slots as requested, provided availability (same as epoch-level but the wavelength for particular timeslots are checked in the WD stage). The matrices in `F' in Fig. \ref{fig_SCHD} show the filling of final wavelength-timeslot resource matrix by the slot-level algorithm for the source and the destination both from the buffer ($i$=1-top) and \textcolor{black}{node} ($i$=2-bottom) requests. After coarse allocation is done, resources are fragmented as shown by the matrices in `F' in the bottom part of slot-level allocation. Hence, after the coarse allocation phase, later iterations of fine allocation aim to utilize these fragments. In the fine allocation phase, each timeslot is re-visited and up to one timeslot is granted per source per iteration. All requests with winning grants are marked to avoid repeating requests in future iterations (also shown by the red line to the buffer). Due to pipelining, the WD module does not have knowledge of the most recent wavelength updates and can still request contradicting wavelengths. In such cases, the requests are rejected and prevented from requesting in the current epoch (shown by red line going to the Invalid Request block). The wavelength-timeslot configuration is updated, registered and sent to the transmitters and receivers of each \textcolor{black}{node} and the timeslot information is sent to the sources.

\subsubsection{Iteration/Buffer Management}
As shown in Fig. \ref{fig_SCHD}, the scheduler has a buffer where failed requests are stored in order to retry in future epochs. Hence, in every epoch, a decision has to be made whether an iteration is to be used for processing requesting from the buffer or the \textcolor{black}{nodes}. The scheduler iterations are managed to cater for the requests from both the buffer and \textcolor{black}{node} every epoch. The size of request residing in the buffer (buffer size) is constantly kept in account. The iteration ratio (buffer:\textcolor{black}{node}) is controlled by $i_{buf}$ as shown in Fig. \ref{fig_SCHD}. Since up to $W$ grants are generated every iteration, the total buffer size is divided by $W$ and multiplied by a buffer coefficient ($R_p$ - usually between 1.6-2.5), which ensures minimal buffer accumulation. In epoch-level scheduling, which is a pointer based scheduling, the pointer stays fixed to a request number in the buffer until a minimum percentage of requests are granted is met. After this, the pointer shifts to the next set of requests. Once the pointer scans through all requests in the buffer or a threshold value is met, whichever happens first, the iterations are used for requests from \textcolor{black}{node}. In slot-level scheduling, the initial iterations deal with requests from the buffer. Once a minimum percentage of buffer requests or a threshold value is reached, the requests from the \textcolor{black}{nodes} are considered.

\begin{figure}[!ht]
    \includegraphics[width=0.45\textwidth]{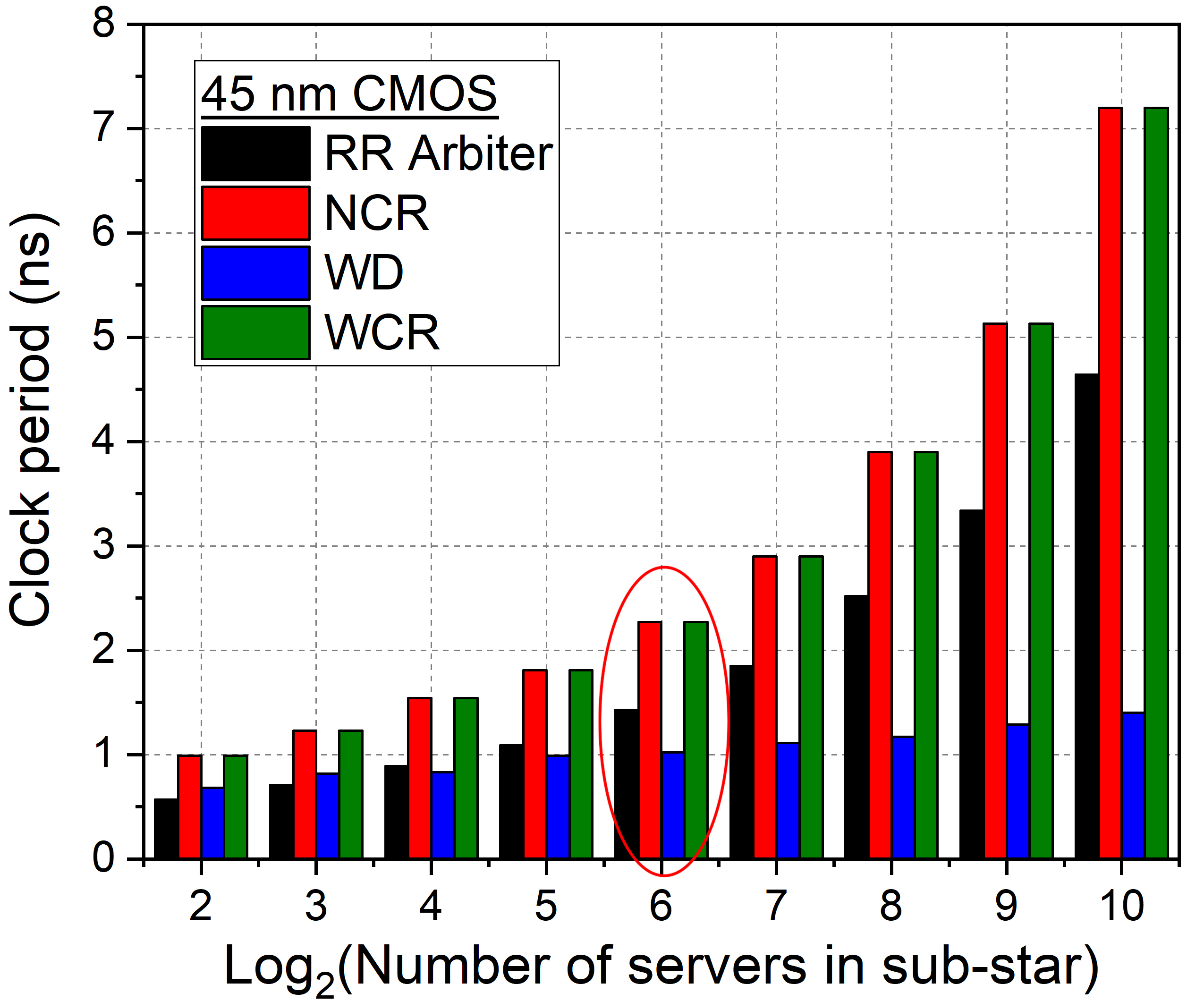}
    \caption{Critical path length of scheduler sub-modules: Node Contention Resolution (NCR), Wavelength Decision (WD) and Wavelength Contention Resolution (WCR) on 45nm CMOS OpenCell library, highlighting N=64 system for this paper.}
    \label{fig_scale}
\vspace{-7pt}
\end{figure}

\begin{figure*}[t]
    \includegraphics[width=1\textwidth]{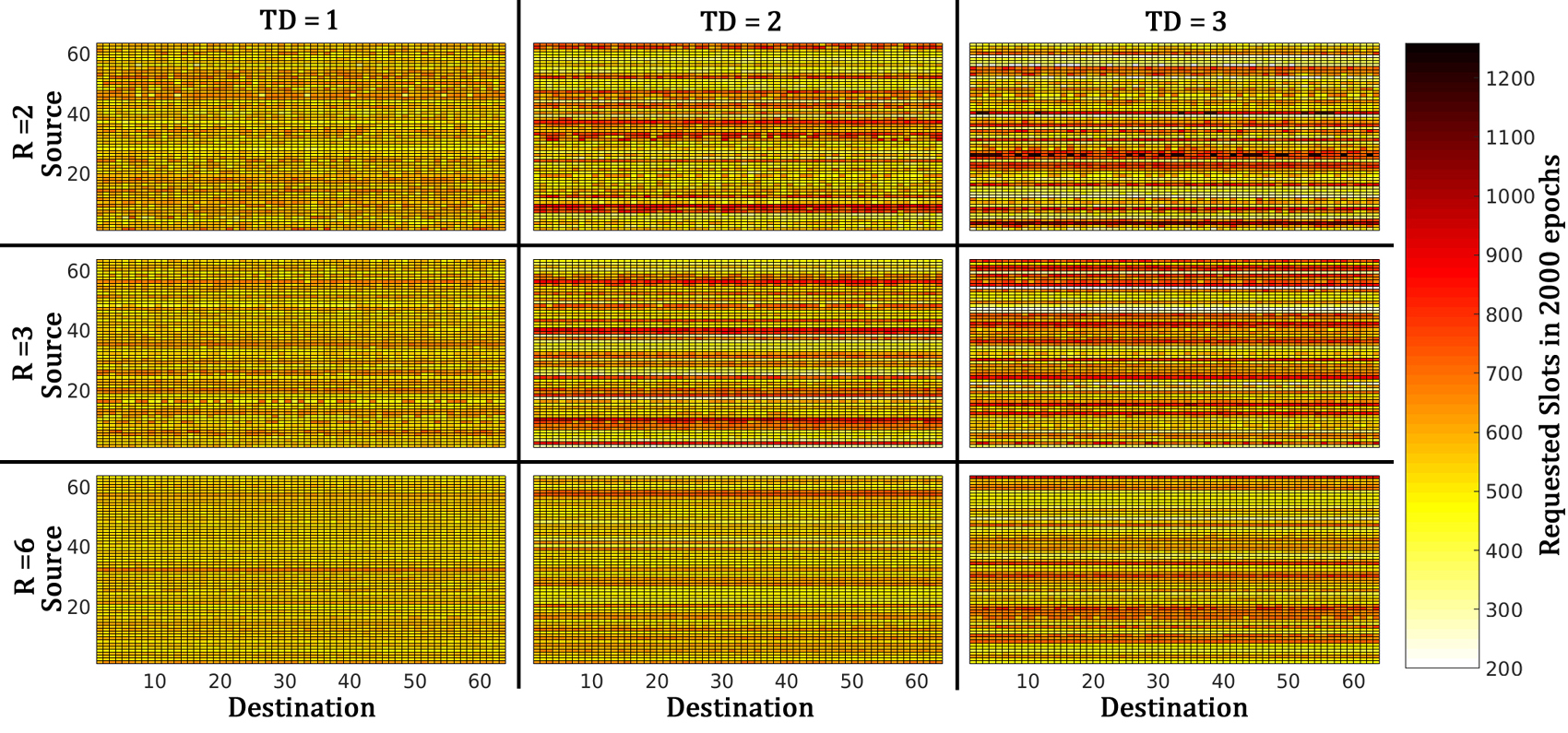}
    \caption{Active \textcolor{black}{nodes}: source-destination demand size for 2000 epochs for 360 ns epoch at 100\% input load.}
    \label{fig_TD}
\vspace{-5pt}
\end{figure*}

\begin{figure*}[!htp]
    \includegraphics[width=1\textwidth]{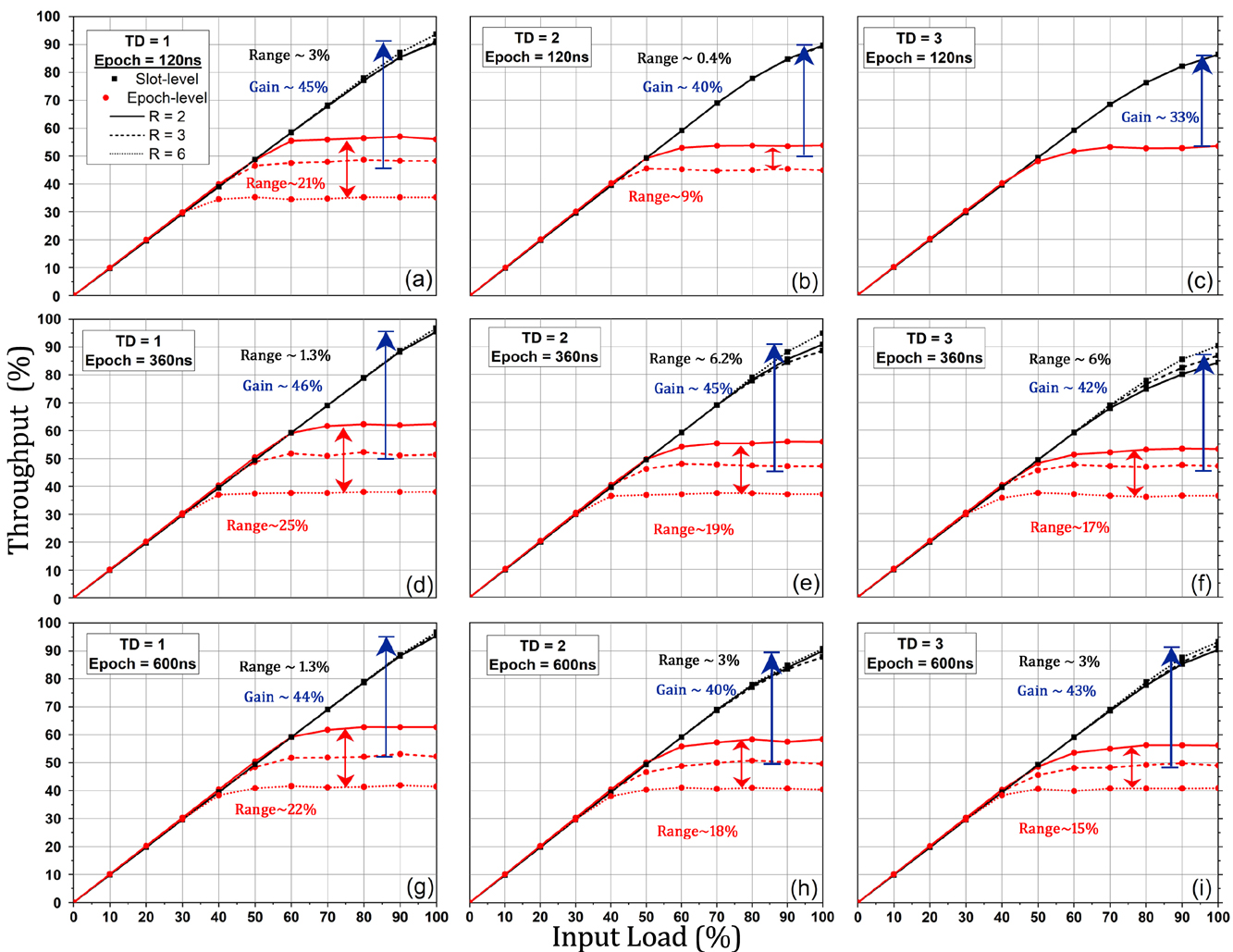}
    \caption{Scheduler throughput vs input load for varying values of R, epoch sizes, TD.}
    \label{fig_thru}
\end{figure*}

\subsection{Implementation}

To demonstrate the functionality of our equivalent software models and hardware design, we simulated the RTL code for the hardware scheduler on Mentor Graphics Modelsim with variable scheduler request traffic inputs from MATLAB. We verified that the outputs were consistent with our software models. Once verified, the scheduler algorithm sub-modules were synthesized on ASIC, using the Synopsys tools and the 45 nm CMOS Nangate Opencell library. We used this technology as it is an open-source, standard-cell library. Figure \ref{fig_scale} demonstrates the scalability of the sub-modules on ASIC as they are scaled to support 1000 ports on 45 nm CMOS technology. The scalability of one $N$:1 round robin arbiter is shown in Fig. \ref{fig_scale}. As parallelism does not significantly affect the critical path delay, $N\times N$-port  and $W\times N$-port  parallel round robin arbiters are used in the NCR and WCR sub-modules respectively. The two arbiter sets in the slot-level allocator are logically identical to the NCR module and are pipelined to have the same size ($N \times N$-bit arbiters). Hence, the scalability of these sub-modules are determined by the scalability of the arbiters. The wavelength decision (WD) stage, however, depends on the scalability of the checker, which uses simple scalable $N$:1 multiplexers to perform the checking (with a critical path of 60~ps per 2:1 multiplexer estimated by the 45nm CMOS OpenCell library). 

The NCR module has the longest of the critical paths, requiring a minimum clock period of 2.3 ns for a $N=64$ \textcolor{black}{node} scheduler. The area consumed by the 1000-\textcolor{black}{node} scheduler ASIC is 52.7 mm$^{2}$ without including the buffers or SERDES in the control plane. The scalability results show that the hardware implementation of these modules is feasible.

%% file: sections/5.results.tex
\section{Performance Analysis}
In this section, we evaluate the performance of the two hardware implementable scheduling algorithms: slot-level and epoch-level allocation. Generating demand traffic with diverse distributions, we study the effect of varying epoch sizes and request loads on throughput, latency, tail latency, transmitter/scheduler buffer size, wavelength usage and energy consumption. Firstly, software models equivalent in functionality to the hardware algorithms were modelled in MATLAB. A request traffic generator is used to feed the scheduling algorithm models with demands and the evaluation of performance is detailed in this section. The parameters and settings that were used to evaluate performance are shown in table \ref{table_1}. As each timeslot can carry up to 250~bytes, a  minimum of 120~ns is investigated, in this paper, to carry the biggest Ethernet packet.

\begin{table}[ht]
\caption{Simulation settings}
\centering
\label{table_1} 
\begin{tabular}{|c|c|c|c|c|c|c|c|c|c|c|}
\hline
\multicolumn{2}{|c|}{\textbf{1. Algorithm}} & \multicolumn{3}{c|}{\textbf{2. Epoch size (ns)}} & \multicolumn{3}{c|}{\textbf{3. R}} & \multicolumn{3}{c|}{\textbf{4. TD}} \\ \hline
Slot & Epoch & 120 & 360 & 600 & 2 & 3 & 6 & 1 & 2 & 3 \\ \hline
\end{tabular}
\end{table}

All possible permutations of 1. Algorithm, 2. Epoch size, 3. Requests per \textcolor{black}{node} per epoch (R) and 4. Traffic Distribution (TD) were used to evaluate the scheduling performance for various input network loads, while each sub-network size supports 64 \textcolor{black}{nodes}/rack (with 64 port schedulers and 64 wavelength channels). However, it is important to note that at 120~ns epoch, there are only 6 timeslots/epoch and hence, some permutations are not possible at high values of TD.

\subsection{Traffic Pattern}

At every epoch, the requests generated by the source are sent to the relevant local scheduler. Each request consists of the destination \textcolor{black}{node}, the number of time-slots required and the origin epoch number. Firstly, to model the demand matrix generation, a uniform random distribution was used to select the destination \textcolor{black}{node} with a probability of $1/N$. Secondly, the average size of each request corresponds to the number of slots available in an epoch divided by the requests per \textcolor{black}{node} per epoch ($S_{avg}=T/R$). Up to $R$ requests are generated by each source per epoch and a Poisson distribution is used to model the inter-arrival rate of requests. All the requests that arrive within the start of epoch are computed, else they are buffered for the next epoch. A uniform distribution of time-slot sizes of requests is used to create the slot traffic distribution (TD) with an average slot size of $S_{avg}$. TD1 corresponds to a single size request, where all requests are of one specific size. In TD2, a total of three size values are allowed (for example, requested timeslots per \textcolor{black}{node}, $S \in 1,2,3$); in TD3, a total of five size values are allowed (for example: $S \in 1,2,3,4,5$). To get a grasp of this double interpolation of uniform random distributions, Fig. \ref{fig_TD} shows the number of slots requested by source-destination pair for a unique TD and R. Focusing one sub-plot, the y-axis shows the source \textcolor{black}{node} number of the source rack and the x-axis shows the destination \textcolor{black}{node} number of the destination rack. The color of the heat-map indicates how many slots are requested over a span of 2000 epochs. The generated traffic is for a 360~ns epoch at 100\% input load. 100\% input load corresponds to all \textcolor{black}{nodes} requesting up to $R$ destinations with $S_{avg}$ timeslots to utilize all resources ($W$ and $T$) available in the sub-network.

As shown in Fig. \ref{fig_TD}, the uniform random request (on both the slot and destination) traffic generation has created a non-uniform pattern. At $R=6$ and TD1, there is the least amount of variation in slot size between source-destination pairs. This is because as the number of request increases, the average timeslot requested is low and all requests are of the same size (TD1). As $R$ reduces to 2, the size of each request increases creating a higher range of variance. As TD changes from 1 to 3, we can see that certain source \textcolor{black}{nodes} become hot in the network, which demand more resources (wavelengths-timeslots) than others. The variations are emphatic as we approach TD3 with low R. This is the type of traffic used to evaluate the performance of the scheduling algorithms.

\begin{figure*}[!hb]
    \includegraphics[width=1\textwidth]{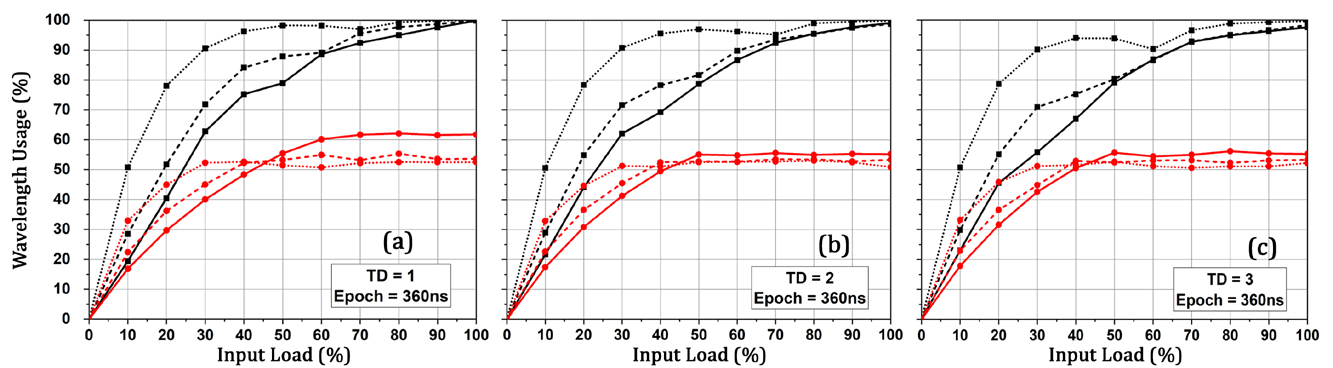}
    \caption{Wavelength usage (out of 64 wavelengths per star coupler) vs Input load for varying R, traffic distributions: (a) TD1, (b) TD2, (c) TD3 in a 360~ns epoch.}
    \label{fig_BP}
\end{figure*}

\begin{figure*}[!htp]
    \includegraphics[width=1\textwidth]{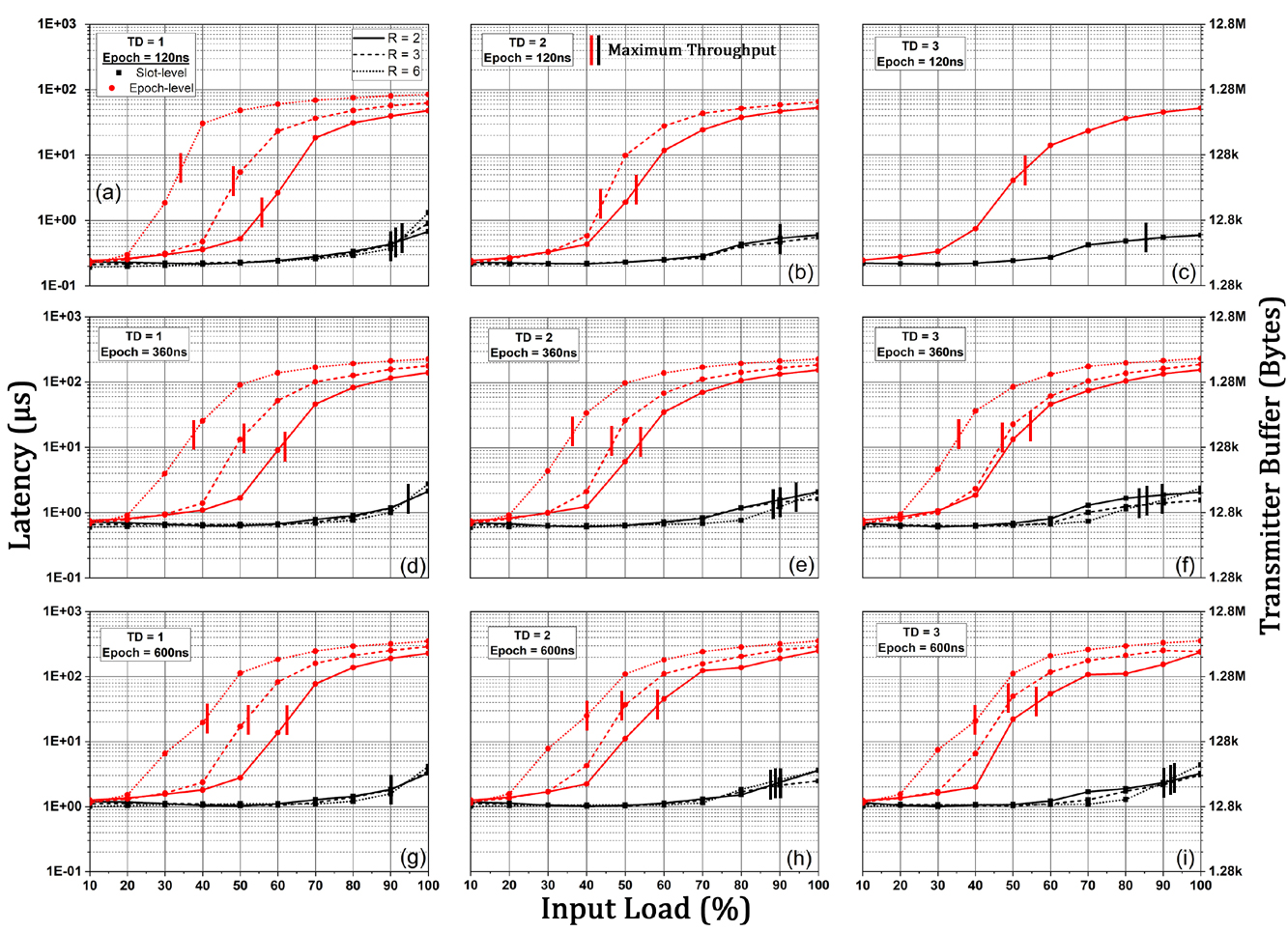}
    \caption{Average end-to-end latency and TX buffer size vs Input load for varying values of R, epoch sizes, TD.}
    \label{fig_Lat}
        \vspace{-7pt}
\end{figure*}

\begin{figure*}[!htp]
    \includegraphics[width=1\textwidth]{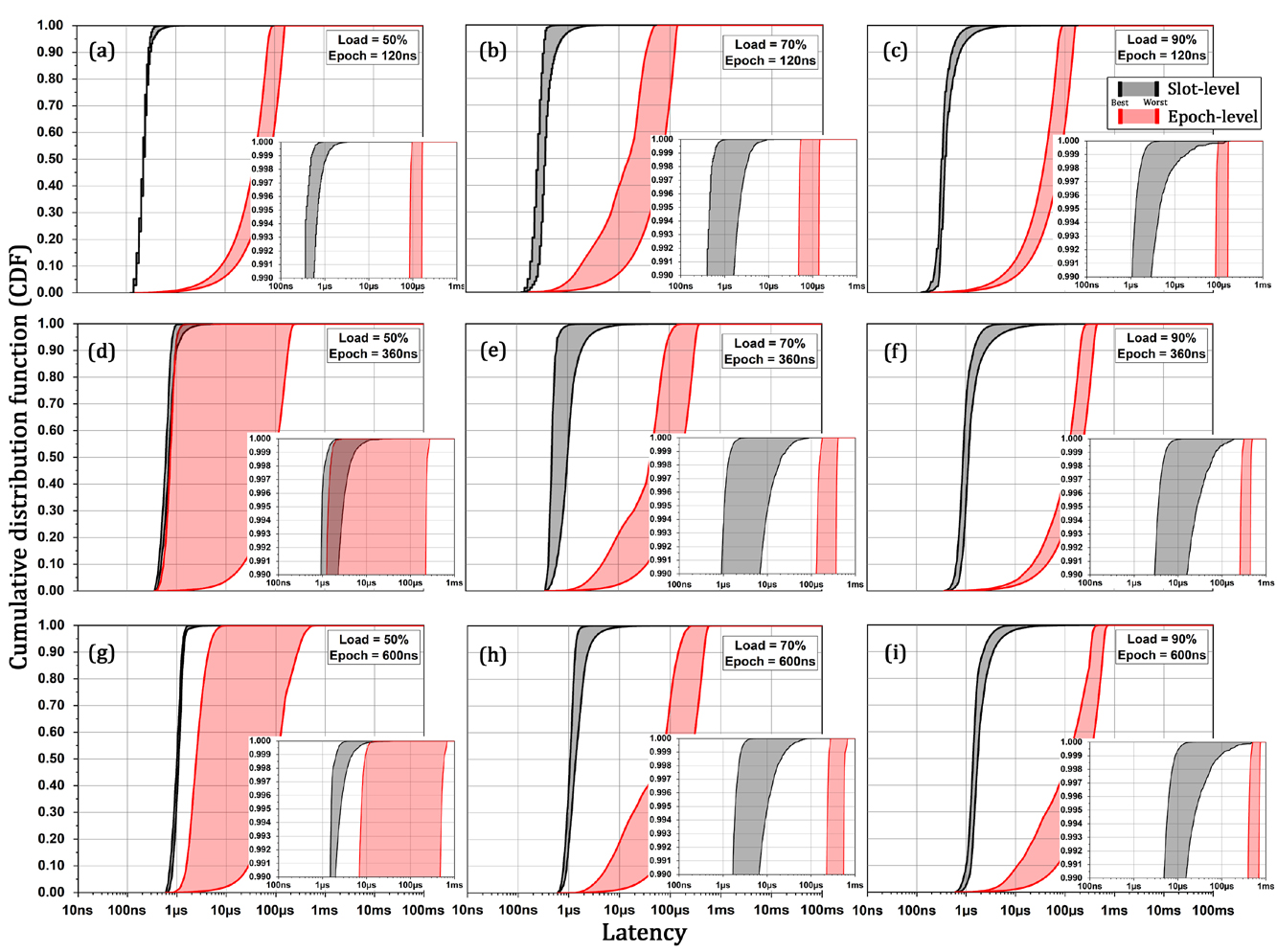}
    \caption{Scheduler latency CDF distribution (inset: tail) for varying values of R, epoch sizes, TD.}
    \label{fig_CDF}
        \vspace{-7pt}
\end{figure*}

\begin{figure*}[t]
    \includegraphics[width=1\textwidth]{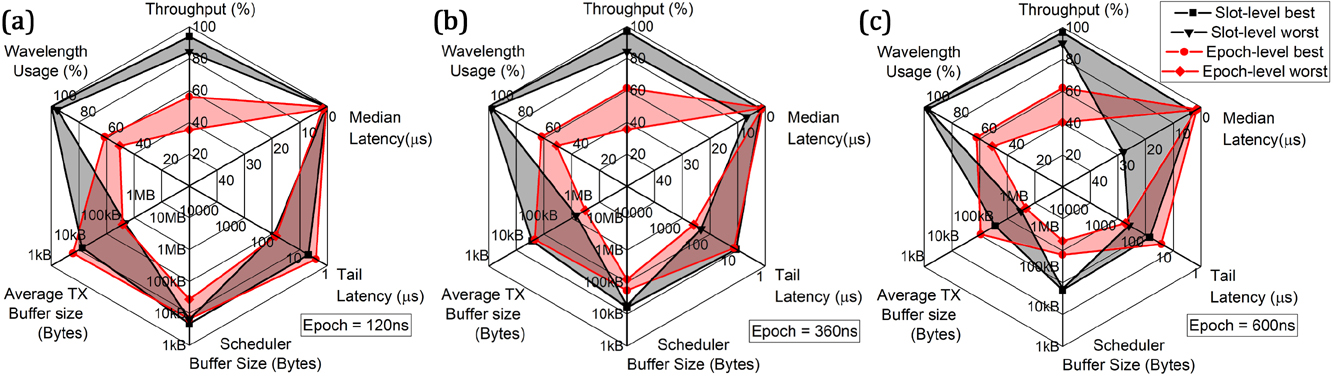}
    \caption{Radar plot - overall performance of epoch-level and slot-level scheduling algorithms at maximum operable load.}
    \label{fig_radar}
    \vspace{-7pt}
\end{figure*}

\begin{figure*}[b]
    \includegraphics[width=1\textwidth]{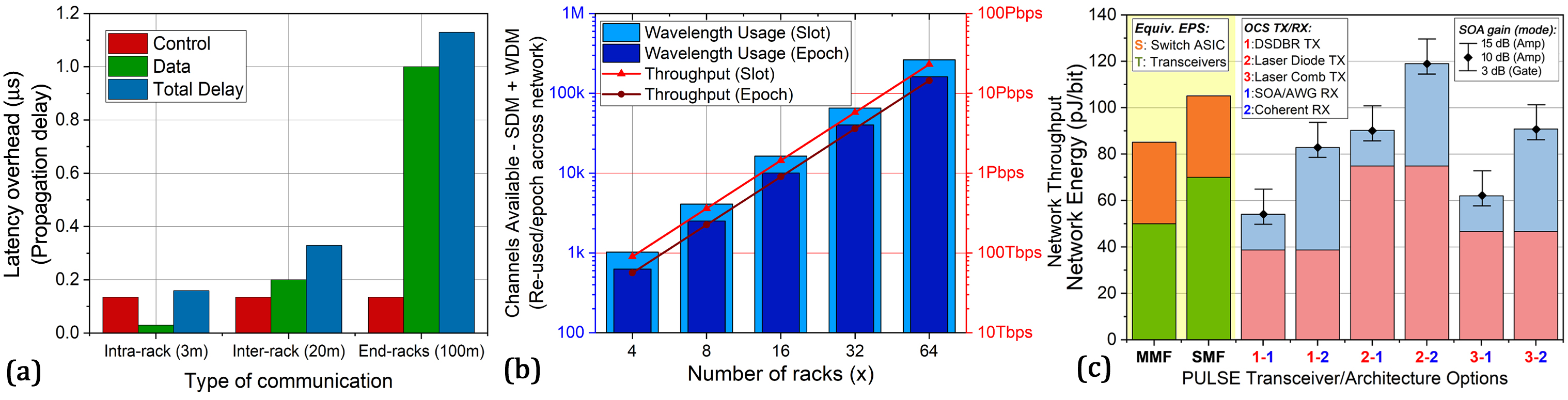}
    \caption{(a) Latency overhead: propagation delay (b) Scaling capacity with x transceivers \textcolor{black}{(c) Network energy consumption.}}
    \label{fig_overhead}
\end{figure*}

\subsection{Throughput Performance}
The throughput of a $N=64$ \textcolor{black}{node} sub-network was evaluated under varying traffic patterns and epoch sizes. The throughput of both the scheduling algorithms (slot-level and epoch-level) were evaluated for increasing input load.  In Fig. \ref{fig_thru}, the effect of changing TD (column) and epoch size (row) on the throughput of both the slot-level and epoch-level scheduling algorithms is shown. Within each sub-plot, the effect of varying $R$ is shown. The values shown take into account the relevant tuning times: 500ps for 20ns timeslot for the slot-level algorithm and 500ps for the entire epoch length for the epoch-level algorithm.

The epoch-level scheduling algorithm in all sub-plots of Fig. \ref{fig_thru} reaches a saturation point at approximate input loads of 60, 50 and 40\% for 2, 3 and 6 requests per \textcolor{black}{node}. In contrast, the slot-level achieves maximum throughput at 100\% input load. The saturation point for the slot-level algorithm is between 85-95\% while the epoch-level saturates between 35-62\%. As evident, the slot-level scheduling algorithm offers an average gain in the matching performance by 32-48\%, compared to epoch-level scheduling for all values of TD and epoch size. 

In the epoch-level scheduling algorithm, each wavelength allocation locks the transmitter and receiver pair to that wavelength for the entire epoch. This increases blocking probability, as future iterations have to work around this \textit{locking}. In contrast, the slot-level allocation has the flexibility of changing the wavelength every timeslot. This flexibility is permitted by the architecture of PULSE's transceiver and network. The variance in throughput in the epoch-level scheduling between $R=2$ and $R=6$ also ranges from 18-30\%. However, this variation is contained in the slot-level scheduling to less than 5\%, showing the algorithm to be more tolerant to variation in request volume. As the number of requests per \textcolor{black}{node} increases, the throughput in both the slot-level and epoch-level algorithm decreases. 

\subsection{Wavelength usage}
Fig. \ref{fig_BP} shows the average percentage of wavelength channels ($W=N=64$ channels) used to achieve the matching for epoch-level and slot-level scheduling algorithms for increasing input loads in a 360~ns epoch. For all values of TD, the maximum wavelength usage in the epoch-level scheduling algorithm is 49-62\%. In contrast, the slot-level scheduling algorithm achieves a maximum wavelength usage of 97-100\%. In both scheduling algorithms, the input load at which the saturation point is reached in the epoch-level algorithm matches the input load at which usage of resources also saturates.  At 100\% input load and low requests/\textcolor{black}{node} ($R=2$), that is low congestion, the saturation point is at 50-60\% load. At the best performance, the epoch-level algorithm is able to use 32-38 wavelength channels every epoch out of 64.  However, at 100\% input load, the slot-level scheduling algorithm saturates very close to 100\%, showcasing a network that utilizes up to 64 wavelength channels.

\subsection{Average Latency and Transmit Buffer}
To measure the latency, each request was marked with a time-stamp when generated. The scheduler has a buffer that stores failed requests in the current epoch to retry the same requests in future epochs. Once successfully and completely (all requested slots) granted, another time-stamp is created to mark when the request will translate into communication in the data plane and reach the destination. The difference in the time-stamps is taken for each request to get the scheduling latency distribution over multiple (2000) epochs. 

Figure \ref{fig_Lat} shows the end-to-end overall average latency taken for packets over 2000 epochs when the network is scheduled using epoch-level and slot-level algorithms for increasing input load, while varying values of requests per \textcolor{black}{node} ($R$ within plot), epoch sizes (row) and traffic distributions $TD$ (column). Each sub-plot also shows the load where maximum throughput is achieved (saturation point).

The propagation delay and tuning overhead are not considered in these measurements; they are discussed at the end of this section. The high throughput achieved by the slot-level algorithm means that many requests are granted without buffering. Hence, the slot-level scheduling algorithm is expected to have a significantly lower latency compared to the epoch-level scheduling, with the minimum latency corresponding to the epoch size. 

The epoch size and $R$ have a greater impact on the latency compared to the TD value. As expected, Fig. \ref{fig_Lat}(a-c) shows that, at 120ns epochs, slot-level scheduling has an average latency of 500-600ns at around 90-94\% saturation load, dependent on $R$, while the epoch-level scheduling algorithm consumes 1-7$\mu$s at 35-55\% saturation load. In the epoch-level scheduling algorithm, higher requests/\textcolor{black}{node} ($R$) have a higher latency, whilst in the slot-level algorithm this variation is quite low. At 100\% load for a 600 ns epoch (worst-case), the significant impact of epoch size on the latency of the scheduling algorithm is clear; the epoch-level scheduling algorithm incurs an average latency as high as 200~$\mu$s, while the slot-level scheduling algorithm incurs an average latency of 4$\mu$s. Hence, the smaller the epoch size, the lower the latency. At 100\% input load, the latency reduction offered by the slot-level compared to epoch-level scheduling is 40-80 times, when using a 120ns epoch. 

Latency has a direct impact on the average transmitter buffer size that is required at the transmitter, as shown by the second y-axis on Fig. \ref{fig_Lat}.  While requests are buffered in the scheduler, the data packets are buffered at the source where they await the grant from the scheduler. An average packet size of 250 bytes is assumed for every time-slot (20ns) delay. Just as latency is reduced by an order of magnitude by the slot-level algorithm relative to the epoch-level scheduling algorithm, the buffer size is also reduced by an order of magnitude. Assuming a 100\% input load, the average buffer size required for a 120ns slot-level scheduling algorithm is 15.36kB, relative to the 1.09MB buffer size demanded by the epoch-level scheduling algorithm. 

\subsection{Latency Distribution}
All requests with a successful grant (contain both timestamps from the time it was generated to the time it was granted) are considered for these latency measurements. For 2000 epochs, considering all $N$-ports and $R$ requests/\textcolor{black}{node}/epoch, a range of successful requests came out with grants. In Fig. \ref{fig_CDF}(a-i), we show the distribution of latency using a cumulative distributed function (CDF) for different epoch lengths ($\in 120, 360, 600$ns) and input loads ($\in 50,70,90$\%).  Each row in Fig. \ref{fig_CDF} shows the scheduling behaviour under different epoch sizes. The curves clearly show a shift to the right and hence, an increase in both the median latency and tail latency. The proportion of this shift in average, median and tail latency is directly proportional to that of the epoch size. When epoch sizes are increased 3 times (360~ns) or 5 times (600~ns), the latency also increases by the same factor. Within each figure, we show the best and worst case for varying values of $R$ and $TD$ to showcase how it affects the latency for both the slot-level and epoch-level algorithm. The inset in each Fig. \ref{fig_CDF}(a-i) shows the behaviour of the tail end of the CDF showing the packets that are buffered for several epochs. This is important to consider because the performance of applications in current data center networks are limited by the tail latency; applications have to wait for hundreds of milliseconds and hang at very high workloads \cite{amazon}. 

The slot-level algorithm can achieve unsaturated throughput above 90\%, while epoch-level algorithms can have their saturation point between 35-62\%. This means that the red curves represented in Fig. \ref{fig_CDF} can only show the best case with many requests still remaining in the buffer beyond the saturation point. The slot-level algorithm shows a two orders of magnitude lower worst-case median and tail latency compared to epoch-level algorithm for all values of input load. In a 120~ns epoch in Fig. \ref{fig_CDF}(a-c), the slot-level algorithm achieves a best-case median and tail latency of 0.4~$\mu$s and a tail latency of 5~$\mu$s respectively, compared to 35~$\mu$s median and 99~$\mu$s achieved by epoch-level algorithm. The worst-case tail in the slot-level algorithm at 90\% input load is at around 120~$\mu$s. This corresponds to latency of almost 1000 epochs and it is mainly caused by few small packets ($<$0.05\%) at high load. 

\subsection{Summary}
In this sub-section, we summarize the results showcased in the previous sections at the saturation input loads; the load at which maximum throughput is reached. In Fig. \ref{fig_radar}(a-c), we show a radar plot showing six different axes: throughput, median latency, tail latency, average scheduler buffer size, average transmitter buffer size and wavelength usage, for 120~ns, 360~ns and 600~ns epoch sizes. For all values of $TD$ and $R$, the best and worst values for identified to make this plot. A curve with a large radius or opening shows that the algorithm has increased efficiency with high throughput and maximal wavelength usage, low latency and buffer size. The best and the worst values are shown in all axes to show the variance or the range of change. Fig. \ref{fig_radar} clearly shows how the slot-level algorithm performs better than the epoch-level scheduling algorithm. The throughput and wavelength usage of slot-level algorithm is above 90\% and tolerant to changes in epoch size compared to epoch-level algorithm 35-62\%. The variance in median latency of slot-level scheduling is lowest for a 120ns epoch, compared to epoch-level scheduling. Maintaining a tolerant buffer size for the scheduler and transmitter, the slot-level scheduling has a better buffer management system.

\subsection{Scalability, Power and Latency overhead}

\begin{table}[!ht]
\caption{Component count, power assumptions per TX/RX}

\label{power}
\centering
\begin{tabular}{lcccccc}
\hline
\textbf{Device} & \textbf{TX1} & \textbf{TX2} & \textbf{TX3} & \textbf{RX1} & \textbf{RX2} & \textbf{\begin{tabular}[c]{@{}c@{}}Power per \\ Unit (mW)\end{tabular}} \\ \hline
\textcolor{black}{\textbf{SOA(on) \cite{soa100}}} & \textcolor{black}{1} & \textcolor{black}{1} & \textcolor{black}{1} & \textcolor{black}{1} & \textcolor{black}{1} & 
\textcolor{black}{405} \\ \hline
\textbf{Comb LD \cite{comb}} & 0 & 0 & 1 & - & - & 1000 \\ \hline
\textbf{Comb-Amp \cite{comb_amp}} & 0 & 0 & 1 & - & - & 1300 \\ \hline
\textbf{LD \cite{vcsel}} & 0 & 64 & 0 & - & - & 80 \\ \hline
\textbf{AWG} & 1 & 1 & 2 & \textcolor{black}{2} & - & - \\ \hline
\textbf{MOD \textcolor{black}{\cite{mod2}}} & 1 & 1 & 1 & - & - & \textcolor{black}{1460} \\ \hline
\textbf{DS-DBR \cite{Dsdbr}} & \textcolor{black}{2} & - & - & \textcolor{black}{-} & \textcolor{black}{2} & 1000 \\ \hline
\textcolor{black}{\textbf{SOA(off) \cite{soa100}}} & \textcolor{black}{1} & \textcolor{black}{63} & \textcolor{black}{63} & \textcolor{black}{63} & \textcolor{black}{1} & \textcolor{black}{8} \\ \hline
\textbf{CO-RX \textcolor{black}{\cite{zhixin}}} & - & - & - & \textcolor{black}{0} & \textcolor{black}{1} & \textcolor{black}{2000} \\ \hline
\textbf{PD \cite{pd}} & - & - & - & \textcolor{black}{1} & \textcolor{black}{0} & 630 \\ \hline
\end{tabular}
    \vspace{-7pt}
\end{table}

The previous sub-sections focused purely on scheduling latency. The modulation, serialization, transceiver latency and the propagation latency were not accounted for. Hence, in this sub-section, we highlight the latency overheads that exist within the network. Fig. \ref{fig_overhead}(a) shows the latency overhead for intra-rack, inter-rack or end-of-rack communication in PULSE, assuming lengths ($L$) of 3m, 20m and 100m  respectively. Each of these links in the data plane corresponds to fiber runs of $L$ meters to the coupler and $L$ meters to the destination. The architecture has the control plane scheduler co-located with the source \textcolor{black}{node} racks within a 3m reach, regardless of where the destination rack resides. The co-location of the scheduler means that the latency overhead of the control plane is a known constant and, since the data plane overhead dominates, the configuration of the network is done long before the data arrives. As shown in Fig. \ref{fig_overhead}(a), a total latency overhead of 0.15$\mu$s, 0.33$\mu$s and 1.12$\mu$s are incurred when communicating intra-rack, inter-rack and end-rack. \textcolor{black}{Integration of large channel bandwidth-dense transceivers as integrated SiP midboard  optics  (MBOs) has been shown in \cite{GZ}, proving  the  feasibility  of  supporting densities of 64 Gbps/mm$^2$ (as of 2014). In 2018, an ASIC switch with in-package optical transceiver ports was demonstrated \cite{inp_trans}. A similar co-packaging with FPGA was reported in \cite{fpga_dem} and demonstrated at \cite{ayar_2019}. Dense SiP integration of transceivers can enable the accommodation of 64 transceivers on a PULSE node.} Fig. \ref{fig_overhead}(b) shows how PULSE can scale with (a)the number of transceivers or racks ($x \in 4, 8, 16, 32, 64$). We show that at $x=64$, we can reuse 0.26M channels and reach a capacity of 23 Pbps. \textcolor{black}{PULSE is a small-scale high bandwidth data center network that scales to support up to 4096 nodes. Novel architectures that can scale to more nodes are being explored.}

\textcolor{black}{With an assumption that each node is equipped with a 100~Gbps transceiver, the cost and the power are normalized against end-to-end 100 Gbps link. In Fig. \ref{fig_overhead}(c), we show the network energy consumption of PULSE. The electronic architecture version of PULSE replaces the star coupler of each sub-star with an EPS switch and two transceivers.} In this architecture, the power consumed when employing MMF transceivers is 85~pJ/bit; however, when using SMF transceiver systems, 105 pJ/bit is consumed per link. State-of-the-art electronic transceivers consume \textcolor{black}{ }\textcolor{black}{3.5~W\cite{finisar_tx}} per 100 GbE port while switch ASICs assume 225~W per 6.4~Tpbs \cite{lee}, their overall contribution to the network is \textcolor{black}{10.5~W/path} in the EPS architecture. The details of the optical components used in PULSE, the power values assumed to estimate the network energy and their references are shown in table III. \textcolor{black}{As shown in table III, an SOA consumes 405 mW (90 mA) if driven to offer an amplification of 10 dB and 8 mW (12.5 mA) if driven in the absorption mode (-20 dB)  \cite{soa100} As shown in Fig. \ref{fig_overhead}(c), the cascaded fast tunable DS-DBR lasers (TX1) requires the least number of devices and consumes lowest power (38 pJ/bit) out of all the proposed transmitter options. The power consumption of the laser diode transmitter (TX2) is as high as 75 pJ/bit because it requires $W$ active laser diodes, one SOA in ON state and $W-1$ in OFF state per transceiver per network. Micro-ring resonator-based laser comb generators (TX3) require laser diode (1W power consumption) \cite{comb} and low-noise, high gain amplifiers (1.3~W power consumption) \cite{comb_amp} to compensate for the coupling, insertion and filtering losses demonstrated in \cite{msr_comb} as well as one SOA in ON state and $W-1$ in OFF state for gating (as per TX2); a total power of 47 pJ/bit. Selecting the DS-DBR transmitter option, (1) fast wavelength tunable filters consumes a power of 54 pJ/bit (5.4 W/port) or (2) coherent receiver technology consumes a power of 82 pJ/bit (8.2 W/port). The range on the bar indicates the power consumed if both the transmitter and receiver SOAs are driven at a different gain level. Coherent receiver is the choice of receiver elected for cost analysis, as the fast wavelength tunable filters (1) require large number of components when integrating on a photonic receiver and (2) scalability increases integration complexity (more wavelengths means more SOAs to be added)}.

\begin{table}[!b]
\caption{\textcolor{black}{Cost estimate of PULSE network compared with equivalent electronic DCN as of 2019}}
\label{cost}
\centering
    
\begin{tabular}{|l|c|c|c|}
\hline
\multicolumn{1}{|c|}{\multirow{2}{*}{\textbf{PULSE Component}}}                                    & \multirow{2}{*}{\textbf{\$/Gbps}} & \multicolumn{2}{c|}{\textbf{\#/path}}                              \\ \cline{3-4} 
\multicolumn{1}{|c|}{}                                                              &                                    & \textbf{EPS} & \textbf{OCS} \\ \hline

\textbf{\begin{tabular}[c]{@{}l@{}}100~GE Transceiver (20m)\cite{trans} \end{tabular}}   & \textcolor{black}{1-}3 & 2                                & -              \\ \hline

\textbf{\begin{tabular}[c]{@{}l@{}}Arista (6.4 Tbps) \cite{aris1,aris2} \end{tabular}}   & \textcolor{black}{6-10} & 1                                & -              \\ \hline
\textbf{\begin{tabular}[c]{@{}l@{}}100~GE Transceiver (Co-Rx low)\end{tabular}}     & \textcolor{black}{4.5}                                 &-      & \multirow{3}{*}{}             \\ \cline{1-3}
\textbf{\begin{tabular}[c]{@{}l@{}}100~GE Transceiver (Co-Rx med)\end{tabular}}     & \textcolor{black}{6}                                 & -                 &{1} \\ \cline{1-3}
\textbf{\begin{tabular}[c]{@{}l@{}}100~GE Transceiver (Co-Rx high)\end{tabular}}     & \textcolor{black}{7.5}                                 & -                 &               \\ \hline
\textbf{Star-Coupler \cite{poncost}}                                                               & 0.04           & -                 & 1           \\ \hline
\textbf{Total (\$/Gbps)}                                                            &                             &\textbf{\textcolor{black}{8-}16}            & \textcolor{black}{\textbf{4.54-7.54}} \\ \hline
\end{tabular}
\vspace{-7pt}
\end{table}

\subsection{\textcolor{black}{Cost estimation and comparison}}
\textcolor{black}{In order to fairly compare PULSE with state-of-the-art electronic networks, the cost of deployments with equivalent bandwidth performance (6.4~Tbps), end-nodes and full bisection bandwidth is evaluated. Cost estimates are normalized to the capacity that each component supports (\$/Gbps); for example, the cost of a 100 GbE transceiver cost is \$3/Gbps \cite{trans} \textcolor{black}{or \$1/Gbps for a 100~GbE multi-mode transceiver \cite{multi}}. From our analysis, the cost of the PULSE electronic network costs \$\textcolor{black}{8-}16/Gbps.  In PULSE, a flat transceiver-based architecture, the normalized cost per path is determined by the transceiver architecture while the transport layer cost is kept to a minimum of 0.04\$/Gbps, assuming a 64-port coupler cost of \$240 based on double the cost of a 64-port splitter in \cite{poncost}. The price of the 100~G coherent receiver transceiver is dependent on the complexity of the receiver architecture and DSP required. We have assumed that the simple to complex coherent receivers are \textcolor{black}{1.5, 2 and 2.5} times the 100 GbE direct detect receiver RUC (relative unit cost) based on the report by ASTRON \cite{co-rx_cost} (as coherent transceivers offer the best energy efficiency - Fig 12(c)). Using \cite{co-rx_cost}, we estimated the RUC for adapting the PULSE transceiver architecture has a worst case RUC is \textcolor{black}{2.01} (2x) compared to 100 GbE transceivers. This shows while employing coherent transceivers, the cost of PULSE is \textcolor{black}{\$4.54-7.54/Gbps}, achieving \textcolor{black}{1.1-3.5} $\times$ cost efficiency compared to equivalent electronic network. \textcolor{black}{PULSE only requires end-node transceiver upgrading during a network upgrade cycle whereas electronic architectures would require both 2$\times$ end-node transceiver plus switch upgrades. Hence, the upgrade cost efficiency over time is low for PULSE OCS relative to the equivelent EPS architecture.} The analysis in \cite{trans} shows that price of transceivers is significantly dropping every year, which would benefit a transceiver-switched architecture like PULSE. Moreover, the report in \cite{co-rx_trend} predicts an annual cost reduction of coherent solutions by 15\% and that beyond 16 WDM channels, they will be more cost effective. The report in \cite{coherent800G} predicts that 800G coherent modules employed in data center networks could approach the cost of \$1/Gbps by 2024.}

%% file: sections/6.related.tex
\section{Related Work}

\begin{table*}[ht]
\caption{The relevance of PULSE with respect to current leading OCS Network Research}
\centering
\label{litrev}
\begin{tabular}{|c|c|c|c|c|c|c|c|c|}
\hline
\multirow{2}{*}{\textbf{Topology}} & \multirow{2}{*}{\textbf{\begin{tabular}[c]{@{}c@{}}Switching\\ Scheme\end{tabular}}} & \multirow{2}{*}{\textbf{TDMA}} & \multirow{2}{*}{\textbf{\begin{tabular}[c]{@{}c@{}}Switch\\ time\end{tabular}}} & \multirow{2}{*}{\textbf{\begin{tabular}[c]{@{}c@{}}TDM Slot\\ Resolution\end{tabular}}} & \multirow{2}{*}{\textbf{\begin{tabular}[c]{@{}c@{}}Min. Circuit \\ Duration\end{tabular}}} & \multirow{2}{*}{\textbf{\begin{tabular}[c]{@{}c@{}}Compute \\ time\end{tabular}}} & \multirow{2}{*}{\textbf{\begin{tabular}[c]{@{}c@{}}Architecture \\ Goal\end{tabular}}} & \multirow{2}{*}{\textbf{Device(s)}} \\
 &  &  &  &  &  &  &  &   \\ \hline
\textbf{Helios \cite{helios}} & \begin{tabular}[c]{@{}c@{}}Hybrid \\ EPS$\backslash$OCS\end{tabular} & No & 12ms & - & O(s) & 15ms & \begin{tabular}[c]{@{}c@{}}Reduce cost, power, \\ switch ports\end{tabular} & \begin{tabular}[c]{@{}c@{}}MEMS\end{tabular} \\ \hline
\textbf{OSA \cite{osa}} & Circuit & No & 14ms & - & O(s) & 290ms & \begin{tabular}[c]{@{}c@{}}Achieve high \\ bisection bandwidth\end{tabular} & WSS, OSM \\ \hline
\textbf{Rotornet \cite{rotor}} & \begin{tabular}[c]{@{}c@{}}Circuit, \\ EPS ToRs\end{tabular} & Yes & 20$\mu$s & O(100$\mu$s) & O(1ms) & - & \begin{tabular}[c]{@{}c@{}}Arbiter less, \\ throughput maximization\end{tabular} & Rotor switch \\ \hline
\textbf{Firefly \cite{firefly}} & \begin{tabular}[c]{@{}c@{}}Circuit, \\ EPS ToRs\end{tabular} & No & 20ms & - & O(s) & 60ms & \begin{tabular}[c]{@{}c@{}}Maximal matching\\  for high throughput\end{tabular} & \begin{tabular}[c]{@{}c@{}}LC, Galvo\\ mirrors\end{tabular} \\ \hline
\textbf{REACToR \cite{reactor}} & \begin{tabular}[c]{@{}c@{}}Hybrid\\ EPS/OCS\end{tabular} & Yes & 30$\mu$s & 185$\mu$s & 1.5ms & O(10$\mu$s) & \begin{tabular}[c]{@{}c@{}}OCS to provide \\ EPS performance\end{tabular} & \begin{tabular}[c]{@{}c@{}}100G OCS, \\ 10G EPS\end{tabular} \\ \hline
\textbf{Mordia \cite{mordia}} & Circuit & Yes & 15$\mu$s & 95$\mu$s & O(100$\mu$s) & O(10$\mu$s) & \begin{tabular}[c]{@{}c@{}}Maximal throughput, \\ faster reconfiguration\end{tabular} & WSS \\ \hline
\textbf{PULSE} & Circuit & Yes & 500ps & 20ns & 40ns & 40ns & \begin{tabular}[c]{@{}c@{}}ns-speed scheduling \\ for ultra-low latency\end{tabular} & \begin{tabular}[c]{@{}c@{}}SOA-based \\ transceivers\end{tabular} \\ \hline
\end{tabular}
\vspace{-7pt}
\end{table*}

In this section, the relevance of PULSE with respect to previously proposed optical switch, network and scheduling solutions for data centers is identified. 
Although the demand for data is ever growing, the pin and ASIC bandwidth of current electronic switches are approaching a limit. Hence, extensive research and development have been invested on expensive high capacity switches. As mentioned in section I, scalable optically switched networks with low deterministic median and tail latencies can be a relative game-changer for data centers in terms of power, cost and latency. However, optical packet switches (OPSs) require optical buffer/queue management, congestion control, casting and complex data exchange protocols. OPSs cannot easily replicate the range of complex methods and functionalities that current electronic
switch ASICs perform. As optical buffers do not exist, optical packet switching schemes either use power hungry optical-to-electronic/electronic-to-optical converters and use electronic buffers or fiber delay lines to support queues. In addition to this, the scalability of the control plane also has an impact on latency of the network, as identified by \cite{Diluc}, emphasizing the need for efficient congestion management. 

Nanosecond speed optical circuit switched (OCS) networks are a perfect solution to this problem as they can keep the latency low and deterministic, while keeping the complexity minimal. OCS networks eliminate queues within the switch, the associative issue of packet loss and the need for addressing. OCS offers flexible adjustment to traffic patterns as circuit establishment can last from a few nanoseconds to several hours. However, the key challenge in designing a fast OCS network is the scalability and speed of the hardware scheduler. 

\subsection{Optical Circuit Switch Solutions}
Recent OCS solutions, switching schemes/techniques, their reconfiguration, computation time and circuit duration, architecture goal and the components they use are shown in table \ref{litrev}. Micro-electro-mechanical system (MEMS) based high capacity optical switches can perform high bandwidth workload off-loading (Glimmer-glass in HELIOS architecture \cite{helios}) and reduce overall electronic switch count, cost and power consumption. However, the slow configuration time of the MEMS-based optical circuit switching, about 27ms (table \ref{litrev}) limits their application to long-lived stable traffic; they need to work in co-ordination with electronic packet switches to cater for diverse types of bursty traffic. A faster single comb driven 2048-port MEMS switch has been proposed that can achieve a switching speed of 20$\mu$s \cite{mems}. However, a MEMS OCS still incurs substantial latency when switching small size data (e.g. 20ns data packets), making them suitable only for long-lived data flows. The OSA optical switch architecture also incurs latency in the order of milliseconds due to the high computation time it requires \cite{osa}. Although the switching time in Rotornet is reduced to tens of microseconds and computation time to zero (cyclic switching), the cycle wide reconfiguration time reduces the quality of service under realistic data center traffic patterns \cite{rotor}. Firefly also suffer from high reconfiguration time requiring mirror switching or guiding \cite{firefly}. REACToR \cite{reactor} and Mordia \cite{mordia} have a fast switching time and support TDMA; however, the long circuit duration of the order of (sub-)milliseconds would result in longer tail latencies. In PULSE, we propose a fast tunable transceiver that coordinates with SOAs to perform switching at 500ps, establishing 20ns timeslots. The circuit is computed for multiple timeslots, 120-600ns in this paper, and a hardware scheduler capable of performing $2.3 \times I_E$ iterations is shown.

\subsection{Optical Scheduler Solutions}
The total switch configuration time is defined by both the data and control plane scalability. A software-defined FPGA based approach to configure small port-count optical switches was demonstrated to take 53 ms \cite{griffin}. Distributed and centralized MAC protocol based heuristics have also been proposed to control optical switches. In the 64-port POTORI (coupler based) switch, a centralized and tailored MAC protocol uses Largest First (LF) and iSLIP scheduling heuristics, which were shown to incur a latency of 10 ms above 80\% workload \cite{potori}. The c-MAC control scheme in the AWG-based petabit switch architecture has an estimated latency of 5 $\mu$s (also for 64-port AWG) for offered network loads above 70\% \cite{peta}. Software based scheduling solutions are orders of magnitude slower than the reconfiguration times of PULSE; hence, the need for hardware based schedulers is clear. The Data center Optical Switch (DOS) architecture uses label extractors in front of AWGRs to identify destinations, resolve contentions and configure tunable wavelength converters (TWCs) \cite{Dos}. DOS  assumes an unrealistic hardware clock speed of 2~GHz for its $N$ input port, $N$ output port arbiter elements. 

In contrast to the above OCS switching technology, PULSE identifies the need for a specially designed hardware architecture that enables fast scheduling. The novel transceiver architecture in PULSE gives greater than 3 orders of magnitude faster switching time at 20ns with minimum circuit duration lasting at 120ns. PULSE scheduler algorithm exploits spatial and temporal parallelism and each iteration can grant up to 64 ($\times S_{avg}$) timeslots in fine (coarse) allocation at 435 MHz. The timeslot level allocation enables PULSE to achieve a highly tolerant throughput at 88-93\%.

\subsection{Synchronization \textcolor{black}{and CDR  locking}}
PULSE requires nanosecond resolution time-slot synchronization (each time-slot is 20 ns). Although practical realisation of this synchronisation is beyond the scope of this paper, it is still a crucial requirement for error-free communication. Prior research has shown optical fiber clock distribution to 1000-ports with jitter less than  12 ps, using mode-locked semiconductor lasers \cite{modelocked}. Reliant on the use of a  dedicated 1 Gbps synchronization plane, the White  Rabbit  project can achieve a clock accuracy better than 1ns and precision better than 50 ps spanning distances over 10km \cite{rabbit}.

\textcolor{black}{Recent practical demonstrations by \cite{kari} have shown clock data recovery (CDR) locking to be achieved in $<625$~ps using \textit{phase caching}. The phase information is required to be updated only once every minute, which makes the CDR settling time a negligible overhead and hence, they are not considered in latency and throughput measurements. This removes preamble needs, although phase has to be updated once per several million epochs. This technique has been demonstrated for a 25~Gbps OOK modulation in real-time and is practical for a 100~Gbps transceiver.}

%% file: sections/7.conclusion.tex
\section{Conclusion}
In this paper, a fast and novel transceiver-based OCS network architecture that enables switching at nanosecond timescales has been proposed. Novel fast hardware schedulers that assign resources (dynamic wavelength and time-slots) in nanoseconds for fast network reconfiguration were designed, implemented and their performance evaluated. The parallel design uses arbiters for selection of requests to grant resources. The implementation of the scheduling algorithm on 45nm CMOS ASIC has a clock period of 2.3ns, equivalent to 435MHz, for a 64-port OCS network. The high clock speed allows the scheduler to perform multiple iterations ($I=\left\lfloor{}E/\text{clk}\right\rfloor-b$) within an epoch (OCS switching rate). We have shown that the scheduler can configure the switch either every epoch (epoch-level) or every time-slot (slot-level). We have also shown that the timeslot level switching scheduling algorithm achieves a high throughput of 88-95\% with a gain of 33-45\% over its epoch-configured counterpart. A low average latency of less than 1.2$\mu$s is achieved by the slot-level scheduling algorithm compared to the 60-80$\mu$s latency incurred by epoch-level scheduling; this is proportional to the average transmitter buffer size (less than 12.8kB at operable load). The median and tail latency are also reduced by 2 orders of magnitude compared to epoch-level scheduling. Tuning at time-slot, rather than at epoch-level, increases wavelength usage to 100\% from almost 60\%. The size of the scheduler buffer is also as low as 1~MB for 2000 epochs, which is almost an order of magnitude lower in slot-level scheduling. The use of cascaded fast tunable DS-DBR lasers at the transmitter and fast coherent receivers for reception help to achieve a low network energy consumption of \textcolor{black}{82}~pJ/bit and \textcolor{black}{with} \textcolor{black}{1.1-3.5} times cost reduction compared to an electronic network equivalent. The PULSE architecture scales to support up to 25.6 Pbps  4096-\textcolor{black}{node} network with 64 racks hosting 64 \textcolor{black}{nodes} each. A latency overhead of 0.15$\mu$s, 0.33$\mu$s and 1.12$\mu$s are shown for intra-rack, inter-rack and end-rack distances due to propagation delay.